\documentclass[acmtog]{acmart}
\usepackage{amsmath}
\usepackage{xcolor}
\usepackage{listings}
\usepackage{algpseudocode}
\usepackage[linesnumbered,ruled,vlined]{algorithm2e}
\AtBeginDocument{%
  \providecommand\BibTeX{{%
    \normalfont B\kern-0.5em{\scshape i\kern-0.25em b}\kern-0.8em\TeX}}}
\setcopyright{none}
\renewcommand\footnotetextcopyrightpermission[1]{} 
\acmSubmissionID{1048}

\definecolor{livid}{HTML}{6a89cc}
\definecolor{aurora}{HTML}{38ada9}
\definecolor{squash}{HTML}{f6b93b}
\begin{document}

\title{Barrier-Augmented Lagrangian for GPU-based Elastodynamic Contact}

\author{Dewen Guo}
\affiliation{%
  \institution{Peking University}
  \city{Beijing}
  \country{China}}
\email{guodewen@pku.edu.cn}
\author{Minchen Li}
\affiliation{%
  \institution{Carnegie Mellon University}
  \city{Pittsburgh}
  \country{United States of America}}
\email{minchernl@gmail.com}
\author{Yin Yang}
\affiliation{%
  \institution{University of Utah}
  \city{Salt Lake City}
  \country{United States of America}}
\email{yangzzzy@gmail.com}
\author{Sheng Li}
\affiliation{%
  \institution{Peking University}
  \city{Beijing}
  \country{China}}
\email{lisheng@pku.edu.cn}
\author{Guoping Wang}
\affiliation{%
  \institution{Peking University}
  \city{Beijing}
  \country{China}}
\email{wgp@pku.edu.cn}

\begin{abstract}

We propose a GPU-based iterative method for accelerated elastodynamic simulation with the log-barrier-based contact model. While Newton's method is a conventional choice for solving the interior-point system, the presence of ill-conditioned log barriers often necessitates a direct solution at each linearized substep and costs substantial storage and computational overhead.
Moreover, constraint sets that vary in each iteration present additional challenges in algorithm convergence.
Our method employs a novel barrier-augmented Lagrangian method to improve system conditioning and solver efficiency by adaptively updating an augmentation constraint sets. This enables the utilization of a scalable, inexact Newton-PCG solver with sparse GPU storage, eliminating the need for direct factorization. We further enhance PCG convergence speed with a domain-decomposed warm start strategy based on an eigenvalue spectrum approximated through our in-time assembly. Demonstrating significant scalability improvements, our method makes simulations previously impractical on 128 GB of CPU memory feasible with only 8 GB of GPU memory and orders-of-magnitude faster. Additionally, our method adeptly handles stiff problems, surpassing the capabilities of existing GPU-based interior-point methods. Our results, validated across various complex collision scenarios involving intricate geometries and large deformations, highlight the exceptional performance of our approach.
\end{abstract}



\begin{teaserfigure}
    \centering
    \includegraphics[width=\linewidth]{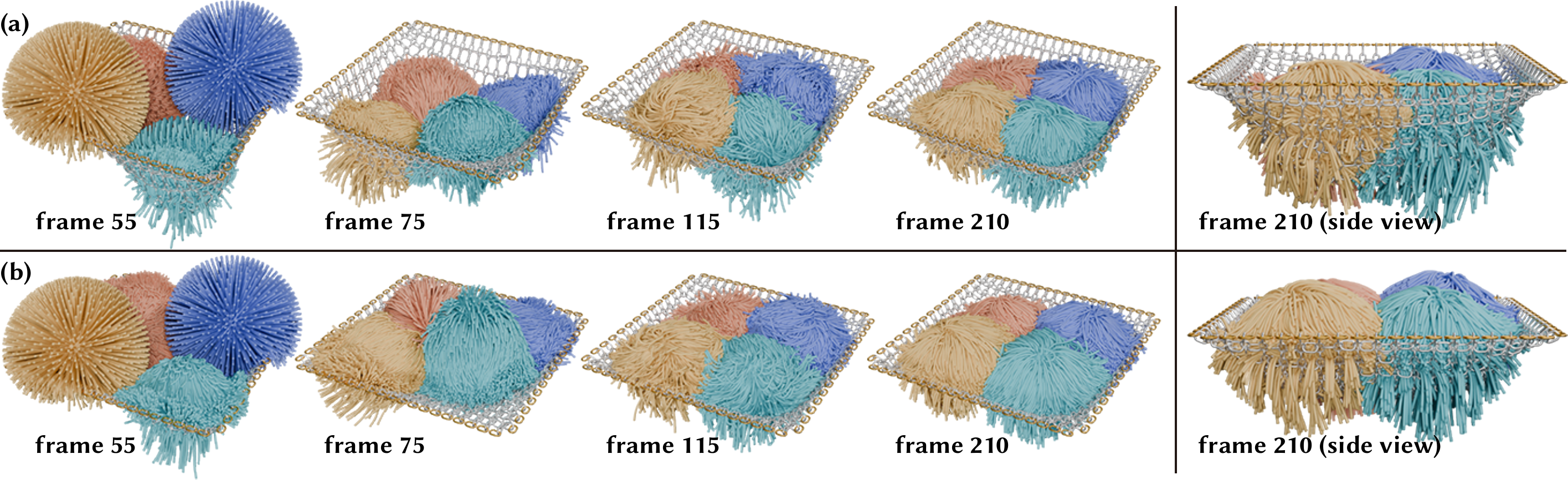}
    \caption{\textbf{Puffer Balls on Nets: Simulations among Heterogeneous Materials.} In this scenario, we simulate the interaction of four puffer balls with a chain-net characterized by different Young's moduli: (a) $E=100$ MPa, and (b) $E=1$ GPa. All puffer balls are modeled using the Neo-Hookean elasticity model with $E=5\times10^{5}$ Pa. Despite the use of high-resolution meshes with over 1.76 million tetrahedra and a large time step size of $1/30$ s, our simulation framework maintains robustness and efficiency. With GPU acceleration implemented, the computation time per frame for scenario (b) is only 427 seconds, without sacrificing accuracy. This represents a notable speedup of $80.1\times$ compared to the IPC \cite{Li2020IPC}, which requires approximately 9.5 hours per frame for the same simulation task.}
    \label{fig:teaser}
\end{teaserfigure}


\maketitle

\section{Introduction}

For robust and accurate simulation of elastodynamics, a common practice in computer graphics is to formulate an optimization problem for an unconditionally stable implicit time integration scheme and then apply the line search method to obtain the solution with guaranteed convergence \cite{gast2015optimization}. The objective function in each time step is called Incremental Potential \cite{kane2000variational}. To achieve fast convergence, search directions are often computed using Newton's method, which solves a 2nd-order approximation of the original problem in each iteration.
A recent contribution named incremental potential contact (IPC)~\cite{Li2020IPC} handles the nonpenetration constraints using a barrier function, enabling robust and accurate contact simulation within the optimization time integration framework.
Unlike complementary programming \cite{anitescu1997formulating}, IPC does not approach the solution by traversing the boundary of the feasible region. Instead, it moves through the interior of the feasible region with infinitely large objective values on the boundary.

Due to the nonlinearity and sharpness of the barrier energy, the direct method, such as Cholesky factorization \cite{chen2008algorithm}, is often incorporated for solving the ill-conditioned linear system in each Newton iteration. Since the factorization will generate a significant number of fill-ins and make the factors much denser, direct solvers are computationally expensive and memory-intensive for large-scale problems. In contrast, iterative methods, such as Conjugate Gradient (CG) or Generalized Minimal RESidual (GMRES), are more storage-friendly and scalable as they only need matrix-vector products to iteratively search for the solution without the need for direct factorization.

However, for iterative linear solvers, convergence is a major concern, which largely depends on the conditioning of the system matrix.
When simulating large deformation or high-speed impacts using IPC, it is not uncommon that the condition number of the Hessian matrix exceeds $10^{10}$, which results from the strong coupling between the highly nonlinear elasticity and the sharp barrier function.
In such situations, iterative methods like CG or GMRES are less effective -- they are either divergent or require a large number of iterations to converge.

Our barrier-augmented Lagrangian method integrates a crucial insight from the performance gains of exterior-point methods: the use of fixed constraint sets until the convergence of subproblems. Exterior-point methods maintain unchanged constraint sets until all current constraints are resolved, a feature that has proven beneficial for practical performance. Traditional methods in contact mechanics, such as impact zone methods \cite{bridson2002robust,harmon2008robust}, face the challenge of requiring restricted step sizes to ensure convergence. To overcome this limitation, mixed exterior-interior point methods \cite{wu2020safe,wang2023fast} have been proposed, utilizing exterior points to guide the solution path while keeping constraints unviolated.
Recently, \citet{lan2023second} introduced a technique for resolving collisions using local CCD within specific local stencils. The efficiency of these methods arises from keeping the constraint sets fixed until subproblems converge, which simplifies the task compared to directly using interior-point methods. The challenge, however, is to integrate this efficiency while maintaining the safety and robustness provided by interior-point methods. In this paper, we adopt the interior-point method as our core model due to its well-established convergence guarantees. Building upon this, we develop an augmented Lagrangian method that incorporates adaptively updated augmentation sets, thus achieving performance improvements comparable to those seen in impact zone and local stencil methods.

\begin{figure}[h]
    \centering
    \includegraphics[width=\linewidth]{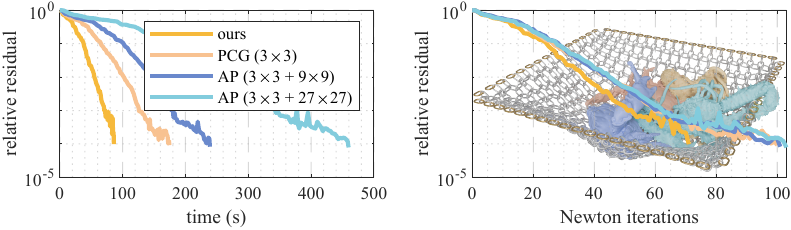}
    \caption{\textbf{Additive Preconditioner (AP) Alone Does Not Yield Performance Improvements.} This is due to the problem's significant nonlinearity caused by varying constraint sets across iterations, which leads to the accurate solutions of the linear subproblems being greatly truncated through line searches. (\#cols = 28,378)}
    \label{fig:schwarz}
\end{figure}

Our method enables smoother application of the Newton-PCG solver for primal problems. To efficiently solve the linear systems, we depart from traditional multigrid or additive preconditioners, which focus on low-frequency error elimination. Instead, we use linear CG as our baseline model and adopt a block-Jacobi warm start by estimating nodal (collision) stiffness. This involves assembling eigenvalues of local contact stencil Hessian matrices into a global diagonal matrix, allowing algebraic decomposition of the simulation domain into stiffness-based groups for separate subsystem solves. Our tests show that additive preconditioners \footnote{The implementation details of additive preconditioner can be found in \autoref{app:as}} can slow down computations, while our method achieves better convergence rate and speed \footnote{The termination criterion is defined as the relative residual, given by $\left\|\mathbf{e}^{[l]}\right\|_2/\left\|\mathbf{e}^{[0]}\right\|_2\le10^{-4}$, where $\mathbf{e}^{[l]}$ represents the residual at the end of the $l$-th Newton iteration.} (see \autoref{fig:schwarz}). Additionally, updating friction constraints per inexact Newton iteration enhances convergence towards a fully-implicit friction model.

Our approach balances storage and computation on the GPU for sparse matrix operations and collision culling using a bounding box hierarchy. The system matrix's sparsity pattern is static without contact events but gains additional non-zero entries when contacts occur. Therefore, storage is divided into element-only and contact stencil components. We developed a specialized Sparse Matrix-Vector Multiplication (SpMV) for our sparse storage, allowing full parallelization on the GPU.

In summary, our main contributions include:
\begin{enumerate}
\item a barrier-augmented Lagrangian method with slack variables that leverages the augmentation sets updated adaptively for improved solver efficiency and system conditioning, along with an adaptive primal-dual optimization scheme for fast convergence (\autoref{sec:barrier_aug_Lag});
\item a GPU-based inexact Newton-PCG solver for the primal problem with fully-implicit friction, featuring algebraically-decomposed block-Jacobi warm start for enhanced performance (\autoref{sec:inexact_newton});
\item scalable GPU strategies for Sparse Matrix-Vector Multiplication (SpMV), collision culling management employing two distinct GPU-constructed linear Bounding Volume Hierarchies (BVH) \cite{lauterbach2009fast}, and floating-point Continuous Collision Detection (CCD) for conservative time-of-impacts (TOIs) (refer to \autoref{sec:scalable_gpu_prog}).
\end{enumerate}
In \autoref{sec:experiments}, we conduct extensive experiments and ablation studies to evaluate our method's efficacy. Our approach shows exceptional robustness and efficiency in handling frictional contact among nonlinear deformable solids, accommodating various material properties and timestep sizes. It maintains consistent performance across different deformation extents and mesh resolutions. Compared to IPC \cite{Li2020IPC}, our method achieves up to a hundredfold speedup, a significant improvement over existing GPU-based iterative methods for complex tasks.

\section{Related work}

\subsection{Elastodynamic Simulation}
Elastodynamic simulation has been a focal point of extensive research within the computer graphics community, spanning several decades since the foundational works by Terzopoulos and Fleischer \shortcite{terzopoulos1987elastically,terzopoulos1988deformable}. Early simulations in this field often applied explicit time integration, which offers simplicity in implementation but imposes limitations on the time step sizes due to numerical instability, particularly when stiff nonlinear elastic materials are involved.
\citet{baraff1998large} proposed the use of implicit time integration to enhance efficiency, laying the groundwork for optimization-based methods.

The robustness of the finite element method (FEM) against element inversion has been improved through invertible SVD \cite{irving2004invertible} and the projected Newton method \cite{teran2005robust}. Recent advancements include inversion-robust strain energies \cite{stomakhin2012energetically,smith2018stable} and analytic eigendecomposition \cite{smith2019analytic,wu2023eigenanalysis}, enhancing the projected Newton method's efficiency. Position-Based Dynamics (PBD) \cite{muller2007position,macklin2016xpbd,muller2005meshless} is popular in video games for its simplicity, stability, and efficiency \cite{fratarcangeli2018parallel}, but it struggles with iteration counts and substepping sizes affecting elastic stiffness. To address this, \citet{bouaziz2014projective} introduced Projective Dynamics (PD), an optimization-based time integration framework that supports various nonlinear elastic materials with ADMM \cite{overby2017admm} and L-BFGS \cite{liu2017quasi}. \citet{WRAPD2021} presents an enhanced ADMM-based algorithm for better convergence and efficiency in geometric optimization, especially with large rotations. \cite{mfem2022} introduces a mixed variational principle for implicit time integration, resulting in a stable solver for different elastic models and timestep sizes. Improvements in the projected Newton method include inexact Newton-PCG \cite{gast2015optimization} and domain-decomposed preconditioning \cite{li2019decomposed}. \citet{macklin2021constraint} extended PBD for singularity-free simulations of the Neo-Hookean model, improving performance without artifacts under stiff conditions. For a detailed review, see \cite{kim2020dynamic}.

With rapid advancements in GPU technology, platforms like CUDA enhance simulation efficiency. \citet{fratarcangeli2016vivace} used Vivace graph coloring to parallelize Gauss-Seidel iterations, while \citet{wang2015chebyshev} accelerated Jacobi and gradient descent methods with the Chebyshev semi-iterative method on the GPU. Error relaxation was further extended to nonlinear descent methods \cite{wang2016descent}. These foundational studies have led to various strategies for accelerating elastic simulations, including numerical methods \cite{wu2022gpu}, reduced-order models \cite{brandt2018hyper}, and geometric considerations \cite{lan2020medial}.

\subsection{Collision Handling}
Contact mechanics are essential for realistic simulations of physical interactions in virtual environments. In computer graphics, this involves addressing friction, deformation, and collision response. Influential works by Baraff and Witkin \cite{baraff1992dynamic,baraff1993issues,baraff1994fast,baraff1994global} have advanced the simulation of rigid and deformable body dynamics. For a comprehensive review of collision handling methods, see \cite{andrews2022contact}; this section focuses on closely related works.

To accurately simulate dynamic interactions between solids, precise collision detection and response are essential. Early studies \cite{provot1997collision,bridson2002robust,harmon2008robust,mirtich1995impulse} used impulses and impact zones with small time steps (e.g., 1/150 s), but did not account for new contact pairs during continuous motion, leading to potential intersection artifacts. Improved continuous collision detection (CCD) methods \cite{zhang2007continuous, brochu2012efficient, tang2010fast, tang2014fast,wang2021large, wang2022fast} have been developed, including a recent root-finding algorithm for higher-order polynomials by \citet{yuksel2022high}. When penetration is acceptable, discrete collision handling can be used, though it does not eliminate all intersections \cite{baraff2003untangling,wicke2006untangling,volino2006resolving}. Recent GPU-accelerated simulators \cite{wu2020safe, tang2018pscc, tang2018cloth, wu2022gpu, wang2023fast} improve cloth and deformable body contact handling but struggle with penetration-free results in large deformations and high-speed motions.
Building upon the advances in collision handling and simulation, \citet{add2020} and \citet{xu2021end} introduce differentiable frameworks aimed at enhancing robotic performance, with the former focusing on a dynamics solver for frictional contact in varied materials, and the latter on a design framework that integrates novel geometric parameterization and a differentiable simulator for contact-rich manipulation tasks.

To enhance robustness, Li et al. introduced Incremental Potential Contact (IPC), an effective method for handling frictional contacts and ensuring intersection- and inversion-free trajectories \cite{Li2020IPC,li2020codimensional,li2023convergent}. IPC has inspired extensive applications: \citet{ferguson2021intersection} used it for rigid body contacts, \citet{lan2022affine} applied it to simulate stiff solids with linear trajectories, and researchers extended it to solid-fluid coupling \cite{xie2023contact}, geometry processing \cite{fang2021guaranteed}, and robot learning \cite{du2023intersection}. For greater efficiency, \citet{lan2021medial} used the medial axis as a contact proxy, and \citet{lan2023second} proposed a stencil-wise second-order descent method on the GPU.
Concurrently, \citet{gipc2024} proposed a Gauss-Newton approximation for contact Hessians to avoid numerical eigendecompositions, enhancing the efficiency of IPC on the GPU.
However, IPC's barrier energy can lead to ill-conditioned systems, complicating iterative linear solvers, and achieving fully-implicit friction requires multiple nonlinear optimizations. We introduce a barrier-augmented Lagrangian method with a slack variable to improve system conditioning, enabling an inexact Newton-PCG method, and achieve faster convergence to fully-implicit friction by updating friction constraints per Newton iteration.

\subsection{Iterative Methods and Multigrid}

Iterative methods like Jacobi or Gauss-Seidel are suitable for GPU implementation but have suboptimal convergence due to their local nature and inability to address global errors. Multigrid methods \cite{brandt1977multi,wang2018parallel} effectively capture and reduce low-frequency errors either algebraically (AMG) or geometrically (GMG) \cite{saad2003iterative,saad1981krylov}. GMG, which uses coarse meshes and transfer operators, offers uniform convergence and optimal complexity, as seen in \citet{xian2019scalable}'s geometric multigrid for projective dynamics. AMG, like NVIDIA AmgX \cite{naumov2015amgx}, avoids hierarchical meshes and provides advanced multigrid and iterative methods \cite{bolz2003sparse}. Schwarz methods \cite{gander2006optimized,dryja1990multilevel,cai1999restricted} divide the domain into subproblems, handled on GPUs using sequential or parallel subspace correction. 
Inspired by these methods, we decompose our simulation domain into groups of DOFs with similar stiffness, solving subdomain systems to warm start the global PCG solve for faster convergence.


\section{Barrier-Augmented Lagrangian Method} \label{sec:barrier_aug_Lag}
\subsection{Problem Definition}
With stacked nodal positions $\mathbf{x}$ and velocities $\mathbf{v}$ after finite element discretization of the simulated solids, we apply backward Euler to time integrate the system from step $t$ to $t+1$:
\begin{subequations}
\begin{align}
    &\mathbf{x}_{t+1} = \mathbf{x}_{t} + h\mathbf{v}_{t+1}, \label{eq:int:x}\\
    &\mathbf{v}_{t+1} = \mathbf{v}_{t} + h\mathbf{M}^{-1}\mathbf{f}_{\blacksquare}\left(\mathbf{x}_{t+1}\right). \label{eq:int:v}
\end{align}
\end{subequations}
Here $\mathbf{M}$ is the lumped mass matrix, and $\mathbf{f}_{\blacksquare}$ is the sum of internal and external forces. Substituting \autoref{eq:int:v} into \autoref{eq:int:x} and including the friction energy $D$ from IPC \cite{Li2020IPC}, the time integration is equivalent to minimizing the Incremental Potential
\begin{equation}
    E\left(\mathbf{x}\right) = \frac{1}{2h^2}\left\|\mathbf{x}-\mathbf{y}\right\|_\mathbf{M}^2 + \Psi\left(\mathbf{x}\right) + D\left(\mathbf{x}\right),
    \nonumber
\end{equation}
obtaining $\mathbf{x}_{t+1}$, followed by the velocity update based on \autoref{eq:int:x}. Here $\Psi$ is the strain energy, $\mathbf{y} = \mathbf{x}_{t}+h\mathbf{v}_{t}+h^2\mathcal{G}$, with $\mathcal{G}$ the gravitational acceleration.
If we handle collision by imposing the distance constraint $d\left(\{\mathbf{x}^{a}\}_{i}\right)>\hat{d}$ between surface primitives, where $\{\mathbf{x}^{a}\}_{i}$ denotes the $i$-th primitive pair in the active primitive set $\{\mathbf{x}^{a}\}$, and $\hat{d}$ is a tiny collision offset that indicates the minimal distance between each pair (which is different from IPC),
the optimization time integration then becomes
\begin{equation}
    \begin{split}
        &\min_{\mathbf{x}}E\left(\mathbf{x},\mathbf{x}_{t},\mathbf{v}_{t}\right),\\
        \text{s.t. }&\Tilde{c}_{i} = \hat{d} - d_{i}\left(\mathbf{x}\right) < 0, i\in\mathcal{A},
    \end{split}
    \label{eq:problem}
\end{equation}
where $d_i\left(\mathbf{x}\right)$ is the simplified notation of $d\left(\{\mathbf{x}^{a}\}_{i}\right)$, and $\mathcal{A}$ denotes the active constraint set. 
Our goal is to establish an iterative method that effectively solves this problem without matrix factorizations. The first challenge we are facing here is the nonlinear and nonsmooth inequality constraint.

\subsection{Formulation}
\subsubsection{Background}
In the context of constrained optimization (\autoref{eq:problem}), an inequality constraint can be transformed into an equality constraint through the introduction of slack variables and multipliers via augmented Lagrangian methods. This conversion process results in \autoref{eq:problem} being transformed into
\begin{equation}
\begin{split}
&\min_{\mathbf{x}} \left\{E\left(\mathbf{x}\right) + \sum\limits_{i\in\mathcal{A}} \mu_{i} c_i\left(\mathbf{x}\right) + \frac{1}{2}\sigma\sum\limits_{i\in\mathcal{A}} c_i^2\left(\mathbf{x}\right)\right\},\\
\text{s.t. }&c_i\left(\mathbf{x}\right) = \hat{d} + s_{i} - d_{i}\left(\mathbf{x}\right) = 0, i\in\mathcal{A},\\
&s_{i}\ge0, i\in\mathcal{A},
\end{split}
\label{eq:ext-penalty}
\end{equation}
where $\mu_i$ are the Lagrangian multipliers, $s_i$ are slack variables and $\sigma$ is the penalty factor.

\subsubsection{Primal and Dual Solve} \label{sec:primal_and_dual_solve}
Observing that $s_i$ does not directly correlate with each other in the primal problem, we eliminate $\mathbf{s}$ by expressing it using $\mathbf{x}$, turning the primal solve to an unconstrained optimization w.r.t. $\mathbf{x}$ only. The primal problem w.r.t. $\mathbf{s}$ can be formulated as
\begin{equation}
    \min\limits_{\mathbf{s}\ge\mathbf{0}} \left\{\sum\limits_{i\in\mathcal{A}}\mu_{i}^{[l]}c_i\left(\mathbf{x}\right)+\frac{1}{2}\sigma^{[l]}\sum\limits_{i\in\mathcal{A}}c_i^2\left(\mathbf{x}\right)\right\}.
    \label{eq:problem-s}
\end{equation}
Next, the slack variables are substituted based on
\begin{equation}
    s_{i}=\max\left\{-\frac{\mu_{i}}{\sigma^{[l]}}-\hat{d}+d_i\left(\mathbf{x}\right),0\right\}, i\in\mathcal{A}.
\end{equation}

\begin{figure}
    \centering
    \includegraphics[width=\linewidth]{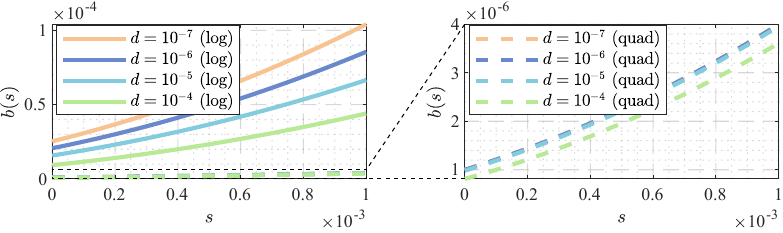}
    \caption{\textbf{Slack Variables} with respect to logarithmic and quadratic penalty, respectively ($\hat{d}=10^{-3}$).}
    \label{fig:barrier-quadratic}
\end{figure}

\subsection{Barrier-Augmented Lagrangian}
The penalty term in \autoref{eq:ext-penalty}, known as the exterior-point quadratic penalty, allows the search outside the feasible region and approaches it from the outside. However, these penalties do not guarantee constraint satisfaction, nor do they ensure a bounded constraint violation in the solution. In contrast, interior-point methods aim to navigate inside the feasible region by introducing log-barrier terms into the objective function. For example, IPC applied a smoothly-clamped $C^2$ barrier function
\begin{equation}
b\left(d_{i}\left(\mathbf{x}\right),\hat{d}\right)=
\begin{cases}
-\left(d_{i}\left(\mathbf{x}\right)-\hat{d}\right)^2\log\left(\frac{d_{i}\left(\mathbf{x}\right)}{\hat{d}}\right), & 0 < d_i\left(\mathbf{x}\right) < \hat{d}, \\
0, & d_i\left(\mathbf{x}\right) \ge \hat{d}
\end{cases}
\label{eq:IPC-barrier}
\end{equation}
to enforce $d_i(\mathbf{x}) > 0$.
Here, we abbreviate $b(d_{i}\left(\mathbf{x}\right),\hat{d})$ as $b_i^{\hat{d}}\left(\mathbf{x}\right)$.
\autoref{eq:ext-penalty} can be regarded as the base model for an exterior-point / impact-zone approach if $c_i\left(\mathbf{x}\right)=s_i-d_i\left(\mathbf{x}\right)=0,i\in\mathcal{A}$, where $\mathcal{A}$  remains unchanged until no constraint violation is detected.
However, previous works \cite{wu2020safe, wang2023fast, lan2023second} demonstrate that interior-point methods can also leverage this concept to enhance performance. This is achieved through adaptively updated constraint sets, safeguarded by regular CCD every few iterations, using either mixed exterior-interior point methods or local CCDs. To guarantee the convergence at large step sizes, we maintain the base formulation as an interior-point method and define an augmentation set $\mathcal{A}'$ to integrate this idea into our method with a variational form.
Specifically, we view $b_i^{\hat{d}+s_i}\left(\mathbf{x}\right)$ as a special penalty function that strives to enforce $d_i(\mathbf{x}) > \hat{d}$ while guaranteeing $d_i(\mathbf{x}) > 0$. We append the penalty term in \autoref{eq:ext-penalty} with $b_i^{\hat{d}}\left(\mathbf{x}\right)$ and obtain the barrier-augmented Lagrangian of IPC:
\begin{equation}
    \mathcal{L}_{\sigma}\left(\mathbf{x},\mathbf{s},\boldsymbol{\mu}\right) = E\left(\mathbf{x}\right) +\sigma\sum\limits_{i\in\mathcal{A}}b_{i}^{\hat{d}}\left(\mathbf{x}\right) + \mathcal{R}\left(\mathbf{x}\right).
    \label{eq:aug-lag}
\end{equation}
Here, $\mathcal{R}\left(\mathbf{x}\right)=\sum\limits_{i\in\mathcal{A}'}\mu_{i}\left(\hat{d} + s_{i} - d_{i} \left(\mathbf{x}\right)\right)+\sigma\sum\limits_{i\in\mathcal{A}'}b_{i}^{\hat{d} + s_{i}}\left(\mathbf{x}\right)$ denotes the augmentation term, where $\mathcal{A}'$ represents the set of constraints for augmentation, constructed based on the observation of the minimum distance (\autoref{alg:al-iter}, lines 3-6).
Here, we exclude the quadratic penalty term since both terms serve the same objective in a general sense, and the logarithmic penalty induces stronger repulsion compared to the quadratic term (see \autoref{fig:barrier-quadratic}).
For the dual problem, we perform the standard first-order update on $\mu$ (\autoref{alg:al-iter}, line 14).

\begin{algorithm}
\caption{The Barrier-Augmented Lagrangian Method.}
\label{alg:al-iter}
\KwData{the variables $\mathbf{x}^{[0]}$, the penalty coefficient $\sigma^{[0]} > 0$, the Lagrange multipliers $\boldsymbol{\mu}^{[0]} = \mathbf{0}$;}
\KwResult{$\mathbf{x}^\ast$}
\For{$l = \left[0,1,2,\cdots\right]$}
{
    update the collision constraint set $\mathcal{A}$;\\

    \If{$\min\limits_{i}d_{i}^{[l+1]}>10^{-2}\hat{d}$}
    {
        $\mathcal{A}'=\emptyset$;
    }

    \ElseIf{$\min\limits_{i}d_{i}^{[l+1]}<\min\limits_{i}d_{i}^{[l]}$ \textbf{or} $\mathcal{A}'==\emptyset$}
    {
        $\mathcal{A}'=\left\{\text{col}_i: d_i<10^{-2}\hat{d},i\in\mathcal{A}\right\}$;
        \Comment{$\text{col}_i$ denotes the collision pair $i$}\\
    }


    $\mathbf{e}^{[l]}=\nabla\mathcal{L}_{\sigma^{[l]}}\left(\mathbf{x}^{[l]},\mathbf{s},\boldsymbol{\mu}^{[l]}\right)$;\\
    $\mathbf{x}^{[l+1]}=\min\limits_{\mathbf{x}}\mathcal{L}_{\sigma^{[l]}}\left(\mathbf{x},\mathbf{s},\boldsymbol{\mu}^{[l]}\right)$;\\

    \If{$\left\|\mathbf{e}^{[l]}\right\|_2/\left\|\mathbf{e}^{[0]}\right\|_2\le10^{-4}$}
    {
        $\mathbf{x}^\ast=\mathbf{x}^{[l+1]}$;\\
        \textbf{break};\Comment{converged}
    }
    \For{$i \in \mathcal{A}'$}
    {
        update $s_i$;\\
        $\mu_i^{[l+1]} = \mu_i^{[l]} + \sigma^{[l]}b_i^{\hat{d}+s_i}$;
    }
    \If{$\min\limits_{i}d_{i}^{[l+1]}<10^{-2}\hat{d}$}
    {
        $\sigma^{[l+1]}=\max\left(1.2\sigma^{[l]},100\sigma^{[0]}\right)$;
    }
}
\end{algorithm}

\subsection{Adaptive Scheduling}
\label{sec:constraints_violation}
\begin{figure}
    \centering
    \includegraphics[width=\linewidth]{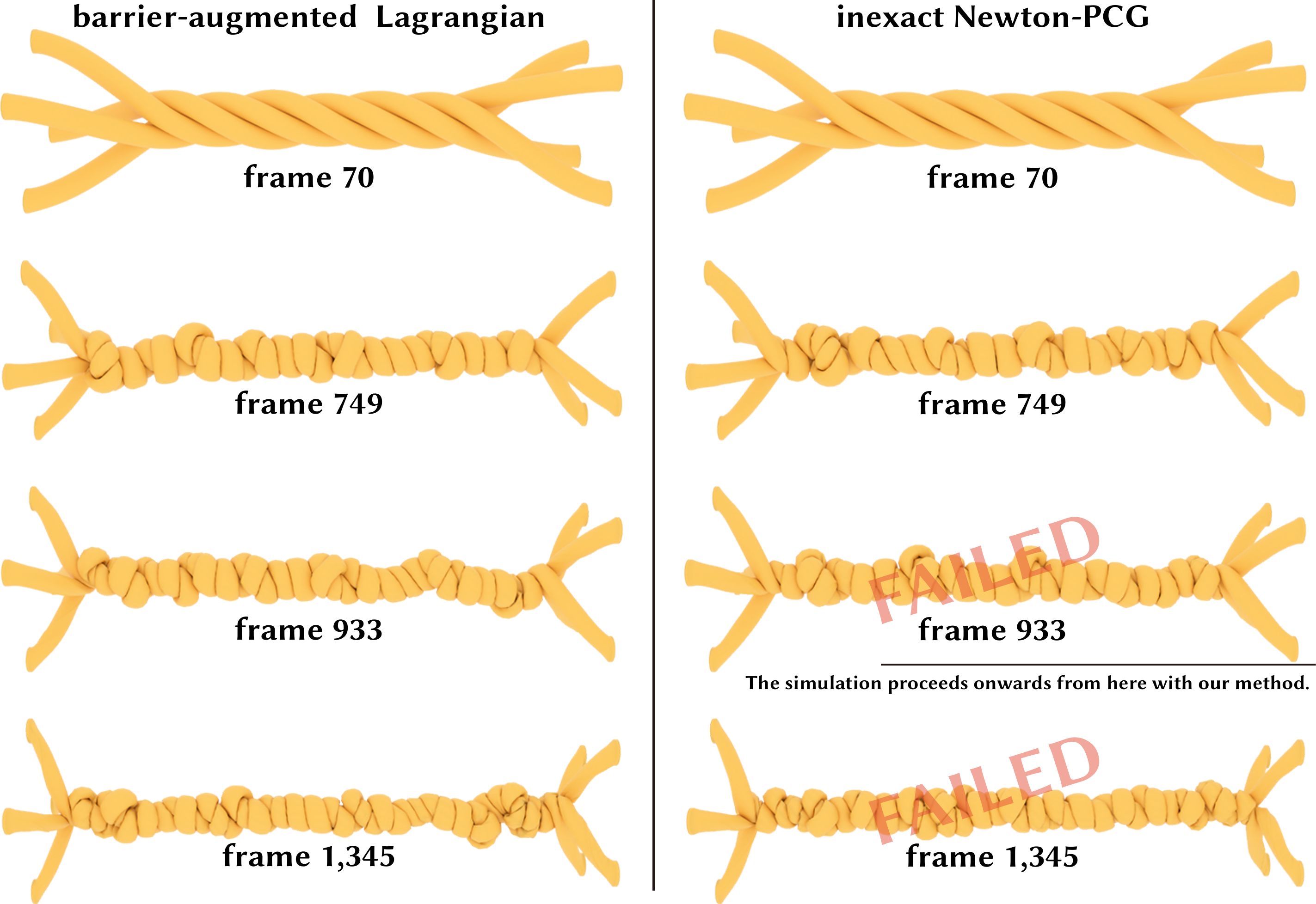}
    \caption{\textbf{Twisting Rods.} This illustration depicts the rigorous stress testing of four stiff rods ($E=10$ MPa), subjected to high-speed torsion from both ends at an angular velocity of 5/12 r/s for over 18 rounds.}
    \label{fig:rods}
\end{figure}

The convergence of primal-dual methods often requires strategic schedules to update the optimization variables and parameters since the optimization is searching for a solution that achieves not only primal optimality but also constraint satisfaction and dual feasibility \cite{nocedal2006numerical}. In this situation, dynamically adjusting the penalty stiffness $\sigma$ is crucial, as it balances the impact of the penalty term against the original objective and the Lagrangian term. A proper balance helps in managing the trade-off between minimizing the objective function and satisfying the constraints. Therefore, we initialize the penalty stiffness, $\sigma^{[0]}$, by solving $\operatorname*{argmin}\limits_{\sigma^{[0]}}\left\|\mathcal{L}_{\sigma^{[0]}}\left(\mathbf{x},\mathbf{s},\mathbf{0}\right)\right\|^{2}$, which gives
\begin{equation}
\sigma^{[0]} = -\frac{\left(\sum\limits_{i\in\mathcal{A}}\nabla b_{i}^{\hat{d} + s_{i}}\left(\mathbf{x}^{[0]}\right)\right)^{\top}\nabla E\left(\mathbf{x}^{[0]}\right)}{\left\|\sum\limits_{i\in\mathcal{A}}\nabla b_{i}^{\hat{d} + s_{i}}\left(\mathbf{x}^{[0]}\right)\right\|_2},
\nonumber
\end{equation}
and then adaptively update it based on the observation of the minimum separation distance by enlarging $\sigma$ to guide the optimization towards stricter constraint satisfaction (\autoref{alg:al-iter}, line 16).
A practical demonstration of our barrier-augmented Lagrangian's efficacy is shown in \autoref{fig:rods}, where we illustrate the twisting of four stiff rods. While the inexact Newton method struggles with convergence at frame 933, our approach effectively overcomes this challenge, enabling continued simulation from the checkpoint (frame 933) where the inexact Newton method stalls.

\section{Inexact Newton-PCG Solver} \label{sec:inexact_newton}

In our augmented Lagrangian framework, we employ an inexact Newton-PCG solver to efficiently solve the primal systems. The critical aspects of our approach include 
the implementation of a warm start strategy (\autoref{sec:precond}) to further enhance PCG efficiency. Moreover, we update the friction constraints per inexact Newton iteration, accelerating convergence towards a more accurate fully-implicit friction model (\autoref{sec:fully_imp_fric}).

\subsection{Fully-Implicit Friction} \label{sec:fully_imp_fric}
\begin{figure}
    \centering
    \includegraphics[width=\linewidth]{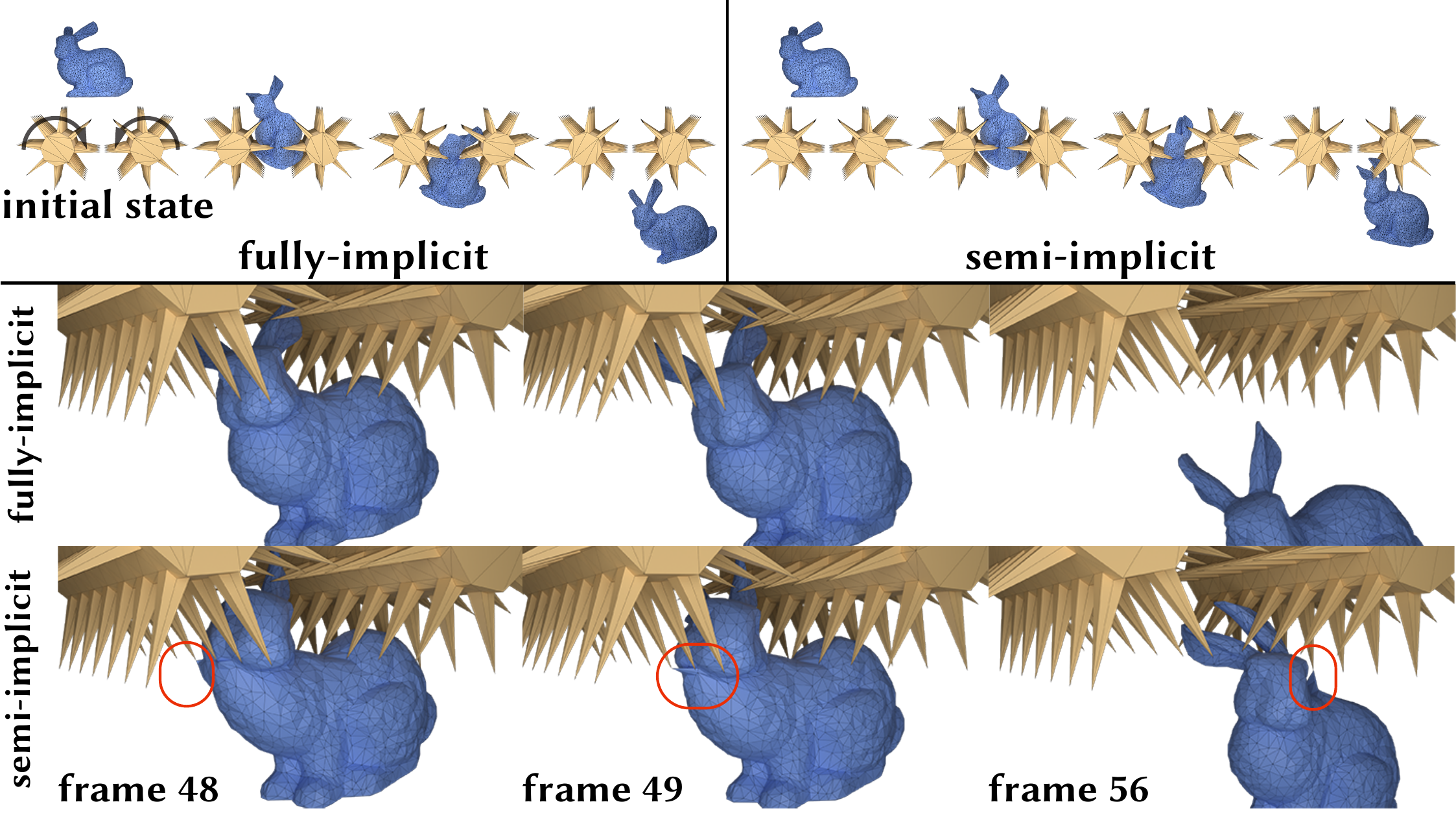}
    \caption{\textbf{The Semi-Implicit Friction May Exhibit Noticeable Sticky Artifacts} when employing a large step size alongside large friction ($\chi=0.9$) in sharp contacts.}
    \label{fig:sticky}
\end{figure}

\begin{figure}
    \centering
    \includegraphics[width=\linewidth]{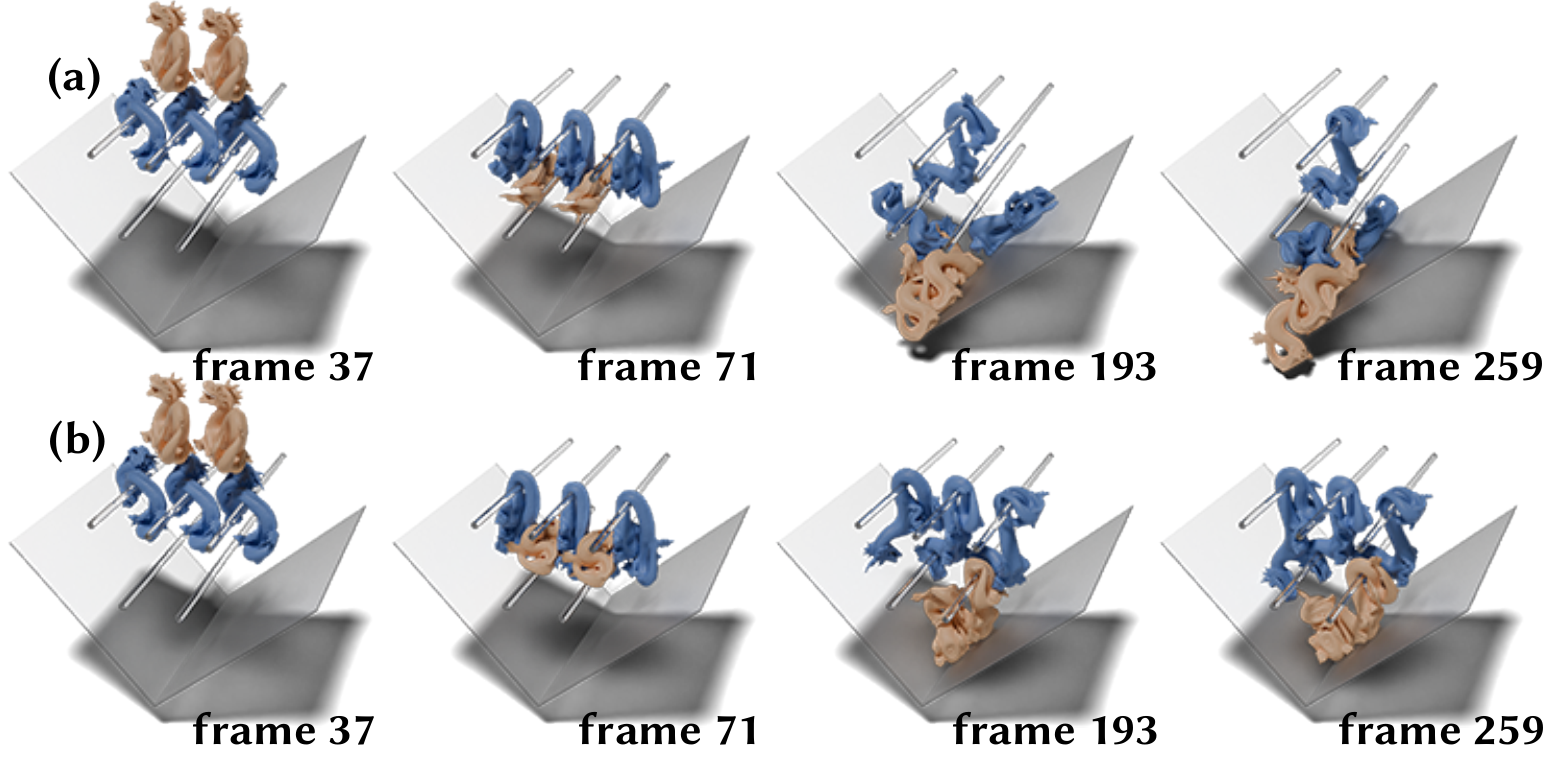}
    \caption{\textbf{Dragons \& Pachinko.} Our fully-implicit friction model accurately captures the dynamics with varying coefficients: (a) $\chi=0.1$ and (b) $\chi=0.3$.}
    \label{fig:dragonsV}
\end{figure}


\begin{figure}
    \centering
    \includegraphics[width=\linewidth]{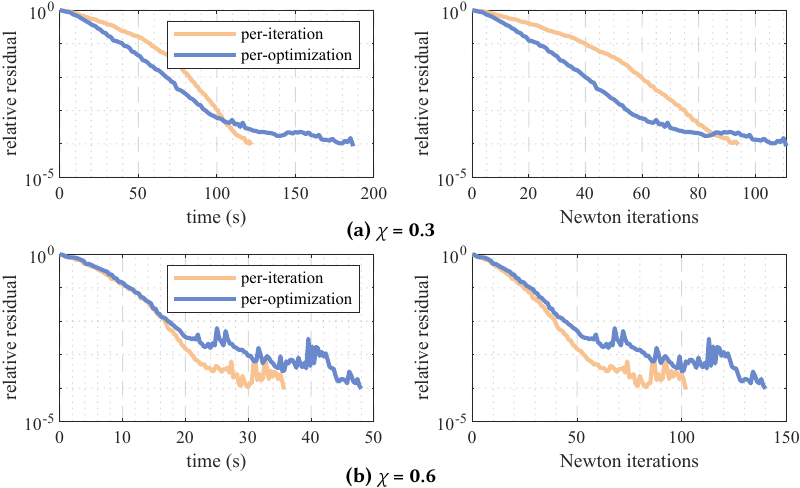}
    \caption{\textbf{Comparison between Per-iteration and Per-optimization Friction Updates.} Our per-iteration updates achieve better performance with fewer inexact Newton iterations. Checkpoints: (a) \autoref{fig:dragonsV}a, frame 193, $\chi=0.3$ (b) \autoref{fig:dragonsV}b, frame 193, $\chi=0.6$.}
    \label{fig:semi_fully}
\end{figure}

As a non-conservative force, friction cannot be directly incorporated into optimization time integration as there is no well-defined potential energy whose gradient will generate friction force.
In IPC \cite{Li2020IPC}, a semi-implicit friction model based on the Maximum Dissipation Principle (MDP) is proposed by discretizing the tangent operator and normal force magnitude of the friction primitive pairs to the last time step, and then an approximated dissipative potential $D$ can be defined as the summation of the energy per friction pair $j$:
\begin{equation}
    \begin{split}
        D_{j}\left(\mathbf{x}^{t+1}\right)=\chi \lambda_{j} f\left(\left\|\mathbf{u}_{j}^{t+1} h\right\|\right), \quad
        \text{where }f\left(y\right) = -\frac{y^3}{3{\epsilon_v}^2h^2}+\frac{y^2}{\epsilon_v h}.
    \end{split}
    \nonumber
\end{equation}
Here, $\chi$ represents the friction coefficient, $\lambda_{j}$ corresponds to the normal force magnitude associated with contact pair $j$, $\mathbf{u}_{j}$ denotes the relative sliding velocity projected onto the lagged contact plane, and $\epsilon_v$ is the threshold in the mollifier $f$.
Although this model ensures guaranteed convergence of the optimization, when dealing with large time steps, the lagged friction constraints may become misaligned with the actual contact scenarios, leading to inaccurate behaviors and even artifacts as demonstated in \autoref{fig:sticky}.
To address this issue, we update the friction constraints per inexact Newton iteration and directly search for the solution with fully-implicit friction.

Specifically, the tangent relative velocity at our Newton iteration $l$ can be computed as
\begin{equation}
    \left(\mathbf{u}_{j}^{t+1}\right)^{[l]} = \left(\mathbf{v}_{r,j}^{t+1}\right)^{[l]} - \frac{\left(\mathbf{v}_{r,j}^{t+1}\right)^{[l]}\cdot\left(\mathbf{n}_{j}^{t+1}\right)^{[l]}}{\left(\mathbf{n}_{j}^{t+1}\right)^{[l]}\cdot\left(\mathbf{n}_{j}^{t+1}\right)^{[l]}}\left(\mathbf{n}_{j}^{t+1}\right)^{[l]}.
    \nonumber
\end{equation}
Here, $\mathbf{n}$ represents the contact normal, and the relative velocity of contact pair $j$ is given by $\left(\mathbf{v}_{r,j}^{t+1}\right)^{[l]}=\frac{1}{h}\boldsymbol{\beta}_{j}^{[l]}\cdot\left(\mathbf{x,}_{j}^{[l]}-\mathbf{x,}_{j}^{t}\right)$, with $\boldsymbol{\beta}_{j}$ being the barycentric coordinates and $\mathbf{x,}_{j}$ representing the subvector of stacked node positions within the contact stencil $j$.
We treat $\lambda$, $\mathbf{n}$, and $\boldsymbol{\beta}$ as constants when differentiating $D$ to compute the semi-implicit friction forces and during the line search, while updating them per inexact Newton iteration to solve for fully-implicit friction. 

In IPC, fully-implicit friction is achieved by updating these friction variables per nonlinear optimization. But convergence is not guaranteed for this sequence of optimizations, which can be interpreted as fixed-point iterations that converge only when starting sufficiently close to the solution (e.g., using a small $h$) \cite{li2022energetically}.

\autoref{fig:dragonsV} showcases five dragons descending into a pachinko-like environment, each experiencing different friction coefficients ($\chi=0.1, 0.3$).
In \autoref{fig:semi_fully}, we compare IPC's per-optimization friction update strategy to our per-iteration strategy within our barrier-augmented Lagrangian framework on the Dragons \& pachinko scenario with larger friction ($\chi=0.3, 0.6$). Our strategy converges to fully-implicit friction with a significant performance gain compared to per-optimization friction updates across divergent $\chi$'s.

\begin{figure}[h]
    \centering
    \includegraphics[width=\linewidth]{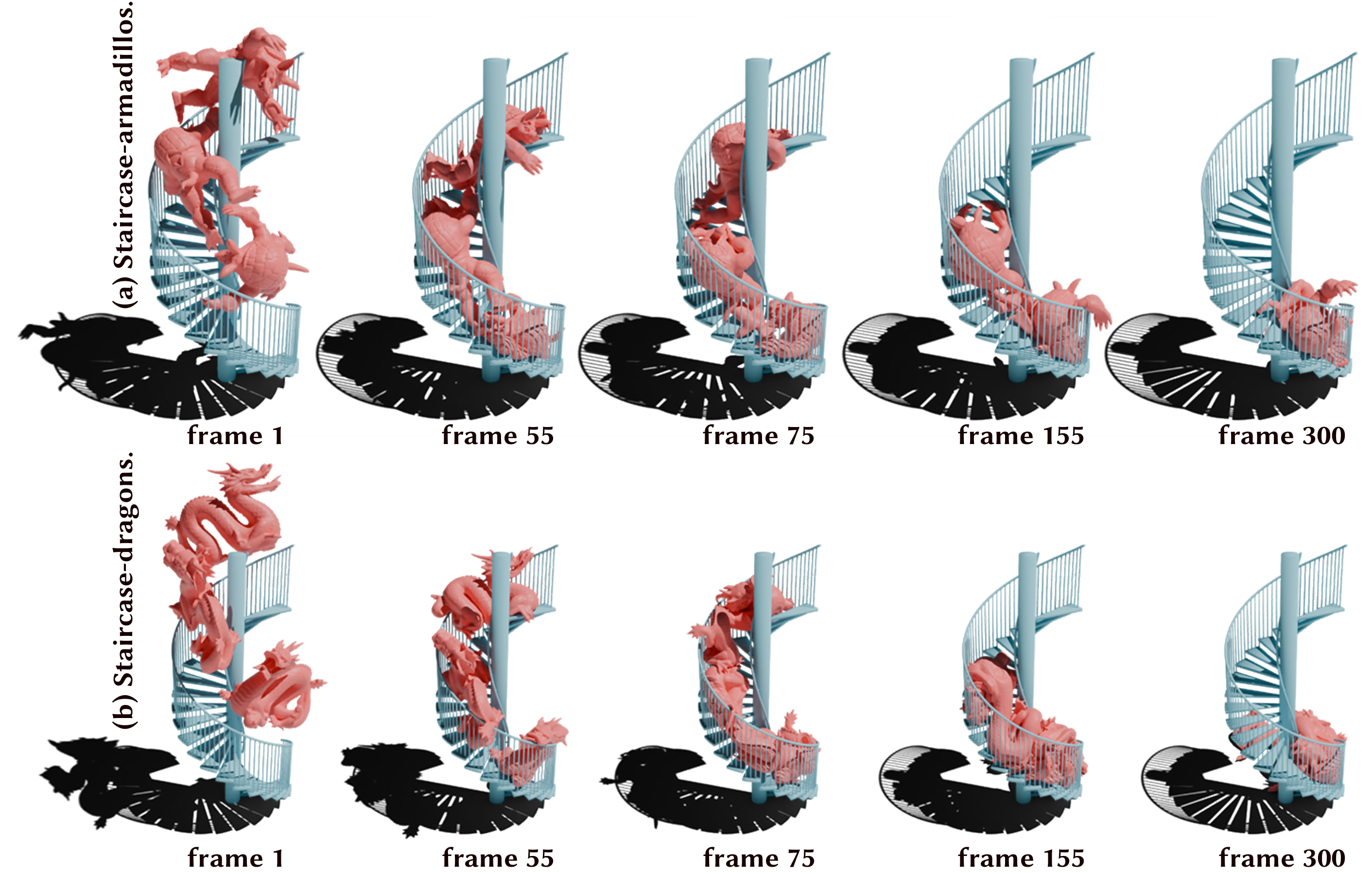}
    \caption{\textbf{Staircase.} Three armadillos (a) or dragons (b) descending along the staircase, engaging in complex collisions. We use these two scenarios to (i) analyze the relationship between node-sorting and sparsity pattern (\autoref{fig:morton-node-sorting}); (ii) demonstrate the superiority of our CCD in ill-tessellated meshes (e.g. the staircase) (\autoref{fig:ccd-fail}); and (iii) compare our block-Jacobi warm start with the GPU-based PCG (\autoref{fig:pcg-cmp}).}
    \label{fig:staircase}
\end{figure}

\begin{figure}[h]
    \centering
    \includegraphics[width=0.7\linewidth]{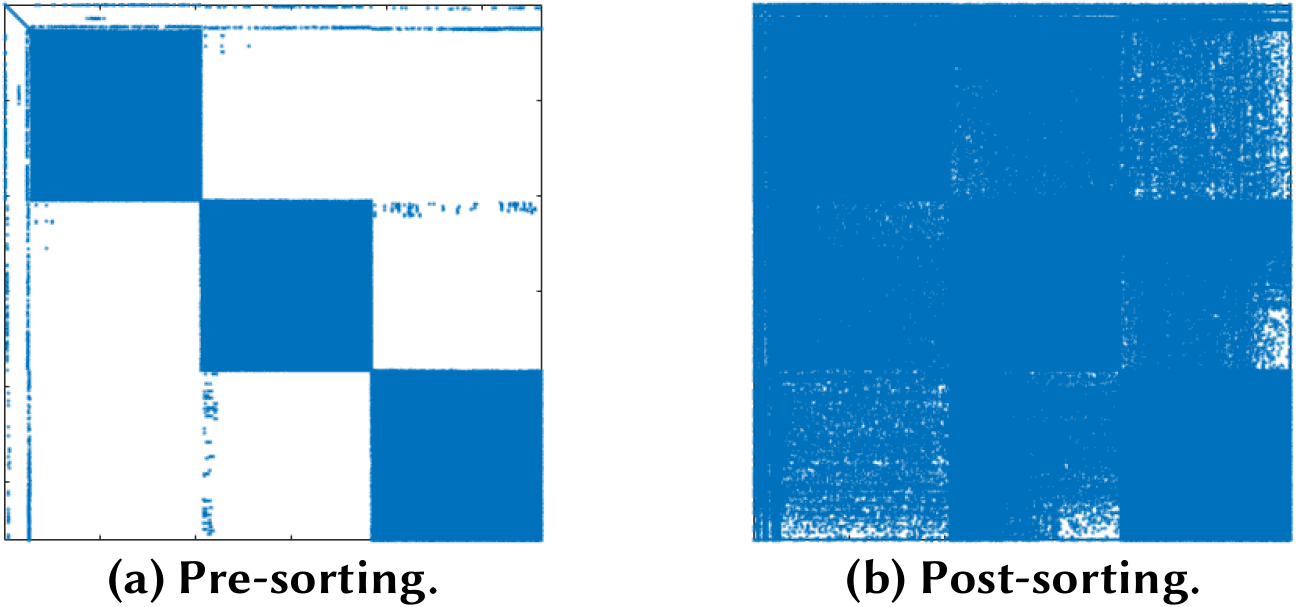}
    \caption{\textbf{Node Sorting with Morton Codes.} Employing Morton codes for node sorting might compromise the sparse linear matrix structure, especially for heterogeneous-tessellated volumetric meshes (scenario: \autoref{fig:staircase}a, frame 75).}
    \label{fig:morton-node-sorting}
\end{figure}

\subsection{Block-Jacobi Warm Start}
\label{sec:precond}

\subsubsection{Discussions on Node Reordering}
In methods for sparse linear systems, matrix factorization is crucial, leading to techniques like matrix reordering or node sorting to reduce fill-ins during factorizations. While current reordering methods focus on graph structures rather than numerical values, they are essential for direct methods relying on full matrix factorization but less effective for iterative methods.
For nonlinear optimizations with changing node graphs, node sorting can be expensive. However, \citet{wu2022gpu} suggest node sorting based on spatial Morton codes to enhance multilevel aggregation, which is particularly effective in cloth simulations due to spatial relations. Still, it may not always apply to volumetric meshes (\autoref{fig:staircase}, \autoref{fig:morton-node-sorting}).
We instead propose a block-Jacobi warm start conveyed implicitly through a novel PCG method, relying solely on SpMV operations and inner products. Utilizing our GPU-based sparse matrix storage (as described in \autoref{sec:spmat}), domain-decomposed computation transforms into a parallel SpMV computation. This process efficiently skips blocks not belonging to the same domain during local matrix-vector multiplications, thus optimizing performance.

\subsubsection{Our Method}
In our barrier-augmented Lagrangian method, although the primal systems become better conditioned compared to projected Newton \cite{Li2020IPC} because of dynamically adjusted adaptive barrier stiffness, a conventional block-Jacobi PCG solver can still converge slowly sometimes. This issue comes from both the drastically different per-node stiffness resulting from the nonuniform deformations and regional self-contact and their large off-diagonal entries in the Hessian matrix. Inspired by \citet{lan2023second}'s success in applying warm start to significantly accelerate solver convergence, we explore a block-Jacobi warm start strategy that separately handles degrees-of-freedom (DOFs) with significantly different stiffness.

To decompose the simulation domain into groups of DOFs with similar stiffness, a straightforward measurement would be the norm of the corresponding row in the Hessian matrix. However, using this norm to measure DOF stiffness may lead to suboptimal decompositions, as each DOF is connected to multiple DOFs by energies with different scales. Recall that to compute descent directions for line search, we have applied stencil-wise eigendecomposition for projecting the local Hessian to the closest symmetric positive semi-definite form. The computed eigenvalues are direct measurements of the stiffness of these connections between the DOFs. We thus take advantage of these intermediate local eigenvalues and assemble them into a global diagonal matrix, where the diagonal entries can then serve as a reasonable estimation of the stiffness per DOF.

Specifically, for each contact stencil, we first compute the local Hessian matrix $\mathcal{H}_c$, and then perform eigendecomposition on $\mathcal{H}_c$ to obtain $\mathcal{H}_c = \mathcal{V}[\boldsymbol{\lambda}]\mathcal{V}^\top$, where $\mathcal{V}$ and $[\boldsymbol{\lambda}]$ are the matrices of eigenvectors and eigenvalues, respectively. 
After eliminating negative eigenvalues and constructing the global Hessian matrix, we proceed to compute the average eigenvalues within each stencil. Subsequently, we form the diagonal matrix $\boldsymbol{\Lambda}$ to approximate the stiffness of the degrees of freedom (DOFs) according to $\boldsymbol{\Lambda} = \sum\limits_{i} \mathbf{S}_i [\boldsymbol{\lambda}_i] \mathbf{S}_i^\top$, 
where $\mathbf{S}_i$ represents the selection matrix of dimensions $3k\times3N$ for stencil $i$. Here, $k$ denotes the number of the nodes in the stencil, and $N$ represents the total number of the nodes in the system.

\begin{figure}[h]
    \centering
    \includegraphics[width=0.9\linewidth]{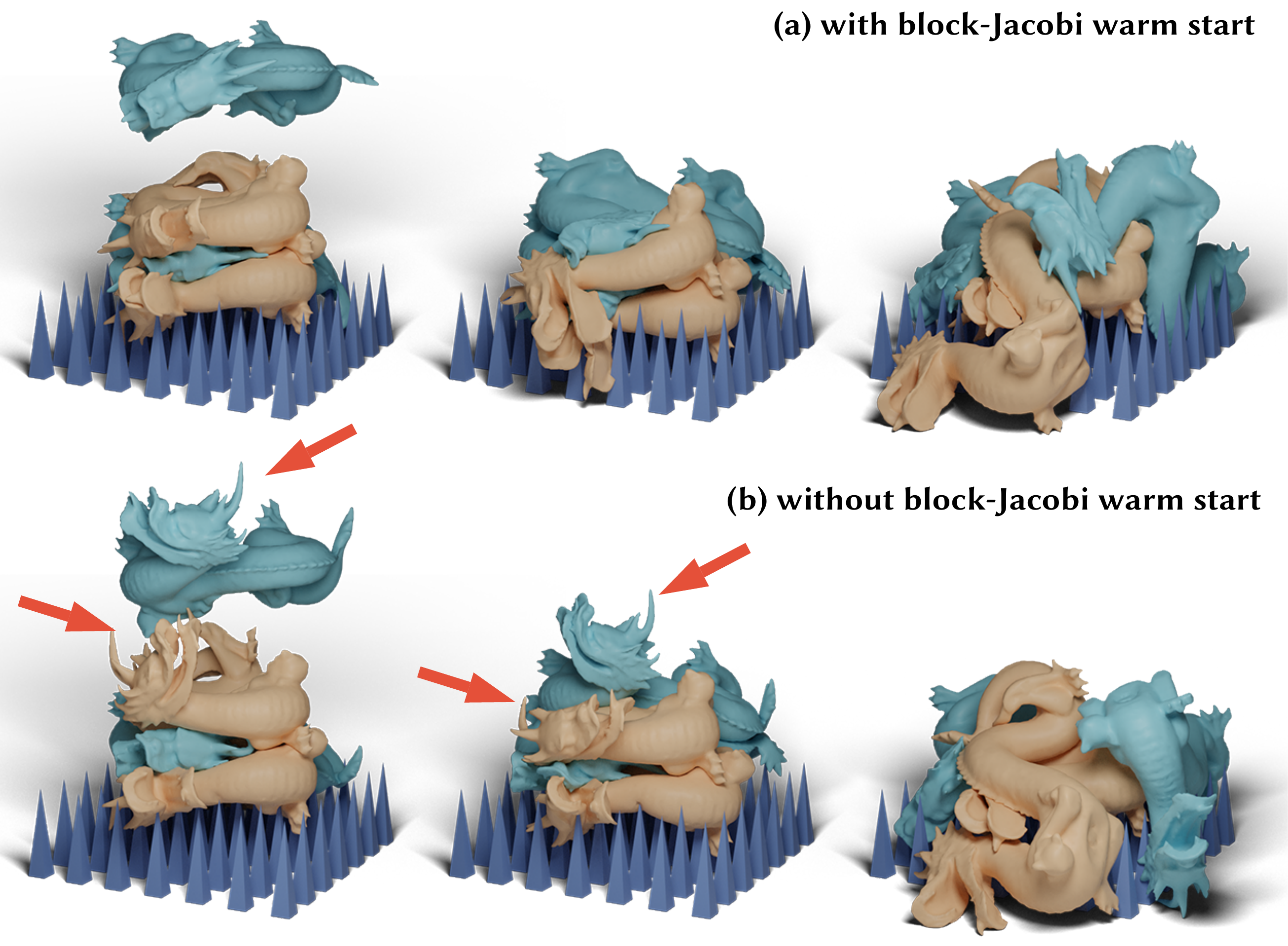}
    \caption{\textbf{Dragons on Spikes.} The absence of our subdomain correction results in noticeable numerical damping artifacts when one of the dragons is in contact with the spikes. Interestingly, even the dragon not directly involved in the collision event is affected (red arrows, leftmost).
    }
    \label{fig:dragons}
\end{figure}

Based on $\boldsymbol{\Lambda}$, we proceed to algebraically decompose our simulation domain, organizing nodes based on their estimated stiffnesses. Let the assembled eigenvalue of node $j$ be denoted as $e_j$, where $j = 1, 2, \cdots, N$. Here, $e_j$ signifies the cumulative sum of assembled eigenvalues spanning from row $3j$ to $3j+2$ in $\boldsymbol{\Lambda}$. We then classify the nodes into groups based on the floor value of their logarithm, $\lfloor\log_{10}(e_j)\rfloor$. These groups form $k$ clusters, where $k\le\max\limits_{j\in N}\lfloor\log_{10}\left(e_j\right)\rfloor+1$ is bounded.

Our block-Jacobi warm start also helps in reducing high-frequency errors more effectively when an early termination of the solver is applied for better efficiency. Please refer to \autoref{app:tolerance} for the PCG tolerance settings. In \autoref{fig:dragons}, four dragons are dropped onto the spikes. The sharp contact forces, together with the resulting nonuniform large deformation in this scenario, lead to an ill-conditioned system. Our block-wise warm start effectively avoids the artifacts near the antenna of the dragons.

\section{Scalable Computation and Efficient Storage on GPU} \label{sec:scalable_gpu_prog}
\subsection{Sparse Matrix Computations} \label{sec:spmat}
\begin{figure}[h]
    \centering
    \includegraphics[width=\linewidth]{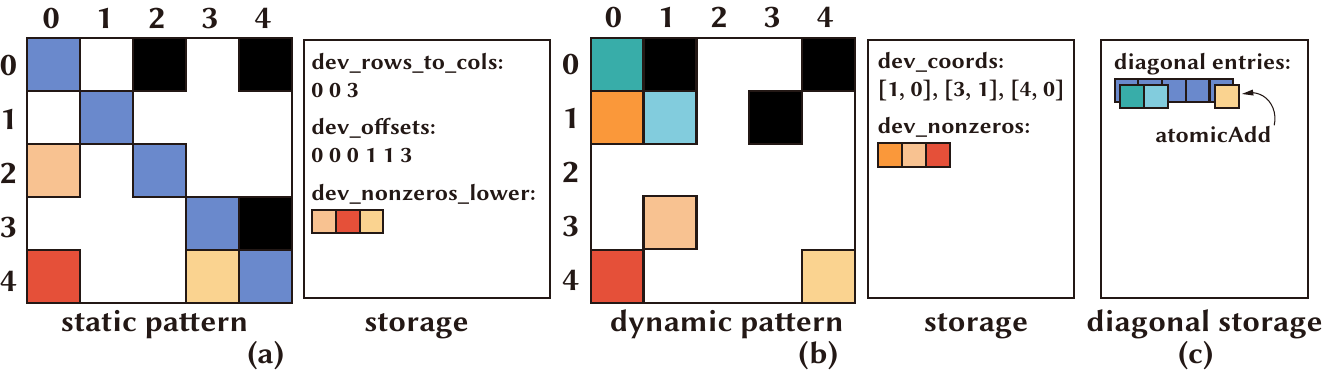}
    \caption{\textbf{Sparse Matrix Storage.} The storage is divided into static (a), dynamic patterns (b), and dense diagonal (c) components, respectively. Since all matrices are symmetric, the upper triangular blocks are disregarded.}
    \label{fig:mat-store}
\end{figure}

In CG iterations, matrix-vector multiplication is the main bottleneck. Thus, efficient storage and sparse matrix-vector multiplication (SpMV) are essential. Since the mesh topology is fixed while the collision stencils often result in different extra connectivity, the Hessian matrices of these two parts can be stored separately to optimize performance.
While matrix-free fashion is often recognized as most effective for CG, it is essential to cache the stencil Hessian during the simulation and optimization process because SpMV will be utilized many times, particularly when the system is ill-conditioned.
We thus exploit the symmetry of the system matrix and store it using three distinct data structures: the block diagonal structure ($\mathbb{D}$) in a dense vector, which is also used for preconditioning; the lower-triangular non-zero entries ($\mathbb{L}$) crafted based on node adjacency; and contact stencil blocks ($\mathbb{C}_i$) stored as pairs of block-coordinate indices along with their corresponding entry values. In this way, SpMV can be parallelized in a block-wise manner on the GPU, employing atomic addition for a map-reduce operation to obtain the overall result. This can be expressed as:
\begin{equation}
    \left(\mathbb{D}+\mathbb{L}+\mathbb{L}^\top+\sum\limits_{i\in\mathcal{A}}\left(\mathbb{C}_i+\mathbb{C}_i^\top\right)\right)\boldsymbol{\upsilon} = \mathbb{D}\boldsymbol{\upsilon}+\mathbb{L}\boldsymbol{\upsilon}+\mathbb{L}^\top\boldsymbol{\upsilon}+\sum\limits_{i\in\mathcal{A}}\mathbb{C}_i\boldsymbol{\upsilon}+\sum\limits_{i\in\mathcal{A}}\mathbb{C}_i^\top\boldsymbol{\upsilon},
    \nonumber
\end{equation}
where $\boldsymbol{\upsilon}$ represents an arbitrary vector in the SpMV operation.
The actual definitions of the data structures are provided in \autoref{app:sp}. For a visual representation, see \autoref{fig:mat-store}.

\begin{figure}[h]
    \centering
    \includegraphics[width=\linewidth]{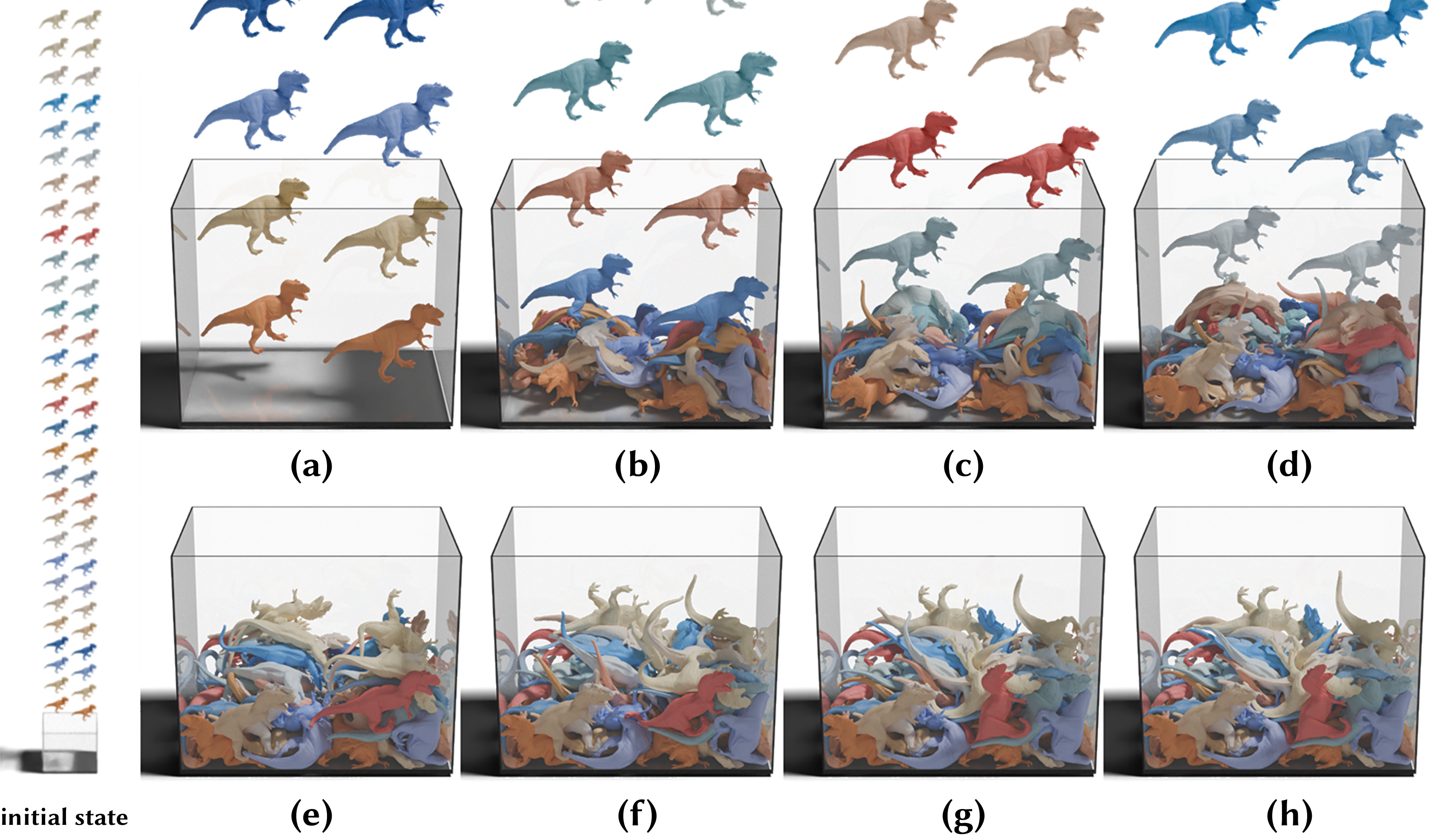}
    \caption{\textbf{T-rex.} In this simulation, 60 T-rexes tumbled into an aquarium, creating a scenario with intense contact interactions. The scene is comprised of over 9 million tetrahedra, yet our method efficiently handles such large-scale problems using just 8GB of GPU memory. Remarkably, the average simulation time for one time step (with a time step size of $h=1/30$ seconds) is only 3 minutes.}
    \label{fig:t-rex-60}
\end{figure}

\subsection{Scalability}
To address scalability concerns, we half the step length of line searches in every Newton iterations until the number of active constraints falls below the specified memory limit. This allows us to effectively process problems on a very large scale. As shown in \autoref{fig:t-rex-60}, we demonstrate the capability of our method by simulating a scene with over 4 million tetrahedra, a scale unmanageable by IPC even with 128 GB of memory. Remarkably, our method successfully processes such extensive scenarios using just 8 GB of memory on the GPU. 
Even with the adoption of AMGCL \cite{Demidov2019} for each Newton solve, this scenario remains impractical and infeasible for IPC, as simulating a single frame requires more than 10 hours.





\subsection{Collision Detection}
\begin{figure}
    \centering
    \includegraphics[width=0.8\linewidth]{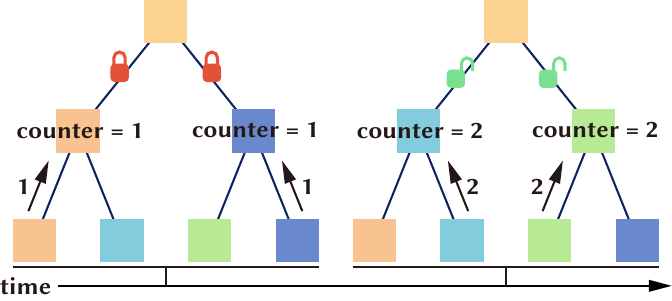}
    \caption{\textbf{Bottom-up Reduction} using \emph{atomicCAS}. Each node has to be visited twice by both of their child nodes without conflict to unlock the gate upstream.}
    \label{fig:atomicCAS}
\end{figure}

\paragraph{Broad Phase}
The collision culling process involves the utilization of a BVH on the GPU for proximity pairs with overlapping axis-aligned bounding boxes (AABBs), such as VF and EE. In our simulation scenario, two distinct trees are generated for triangles and edges, respectively, rather than exclusively for triangles.
This approach is chosen because when AABBs for two triangles overlap, additional CCD tests for 15 VF or EE pairs become necessary. These tests are executed sequentially within the same thread.
\textcolor{blue}{
}
Our base model for BVH is the linear BVH \cite{lauterbach2009fast}, renowned for its exceptional efficiency in real-time construction on the GPU. Linear BVH simplifies the generation of node hierarchies into a sorting problem, where primitives, specifically triangles and edges in our setting, are ordered along a space-filling curve with a Morton code \cite{vinkler2017extended,park2016analysis} assigned to each primitive. We employ a 64-bit encoding system for the position of each primitive, allocating 32 bits for the Morton code and an additional 32 bits for safety considerations in case multiple primitives share the same code after assignment.
Subsequently, the array of 64-bit codes undergoes sorting into lexicographical order using the radix sort. Following this, a binary radix tree is constructed on the GPU, serving as the skeleton of our BVH.
Concurrently, the AABB for each primitive is computed. The AABB undergoes continuous updating for each node through a bottom-up reduction on the tree, a process facilitated by the atomic operation of compare-and-swap (\texttt{atomicCAS}). Notably, each node must be visited twice without conflict before granting access to the upper level, as illustrated in Figure \ref{fig:atomicCAS}.

\paragraph{Narrow Phase}
Efficiently implementing a fast and accurate floating-point CCD on the GPU presents a formidable challenge. To address this issue, we employ a polynomial root finder \cite{yuksel2022high} to solve cubic equations.
A cubic function may have up to three roots. Once the first root $x^{\ast}$ is found, we employ a more efficient strategy known as deflation. The cubic function can be deflated to a product of a linear function and a quadratic form using the expression $ax^3+bx^2+cx+d = \left(x-x^{\ast}\right)\left(Ax^2+Bx+c\right)$, where $A=a$, $B = b+Ax^{\ast}$, and $C=c+Bx^{\ast}=d$.
The cubic polynomial root finder utilizes a Newton-bisection approach. To enhance its robustness, we adopt a conservative modification by setting the interval from $\left[0, 1\right]$ to $\left[-\varepsilon, 1+\varepsilon\right]$.
After solving the cubic equation, we obtain the time-of-coplanar (TOC) for the vertices of potential collision pairs. Subsequently, we calculate the distance of VF / EE candidates, or their degenerate cases, denoting this distance as $d_{\text{TOC}}$. The activation condition is set as $d_{\text{TOC}} < \varepsilon + \hat{d}$ for safety considerations, where $\varepsilon = 10^{-12}$ in our setting.
\begin{figure}[h]
    \centering
    \includegraphics[width=\linewidth]{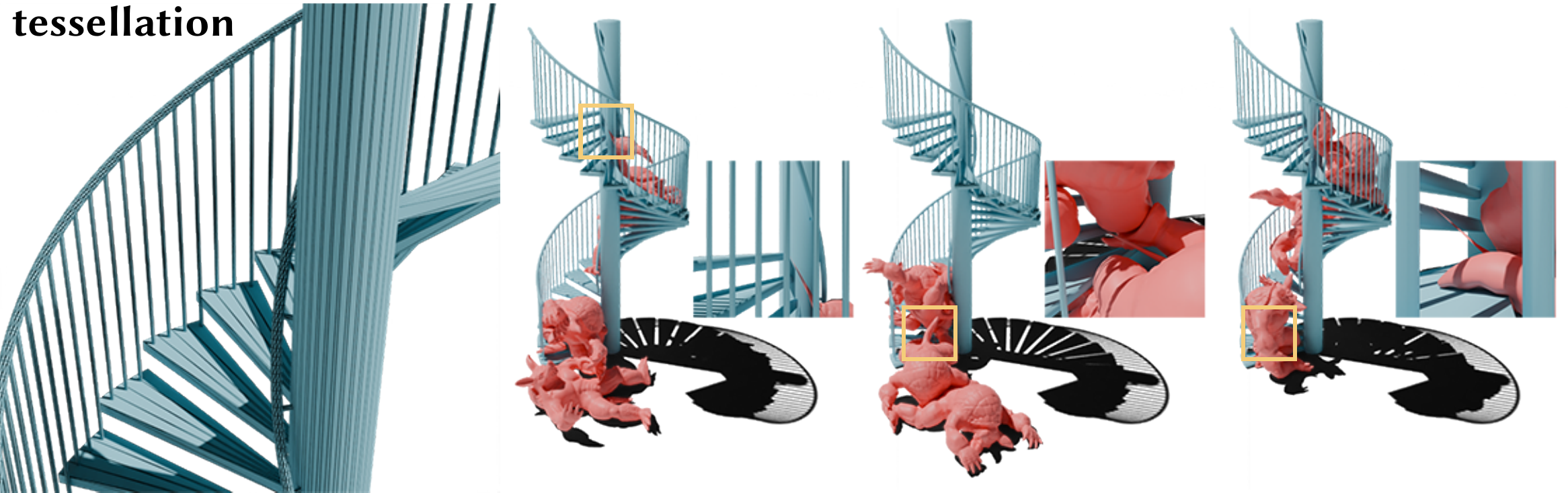}
    \caption{\textbf{The Failure Cases of the Cubic CCD without Optimized TOI} in several runs, where the staircase mesh is ill-tessellated.}
    \label{fig:ccd-fail}
\end{figure}
\begin{figure}[h]
    \centering
    \includegraphics[width=\linewidth]{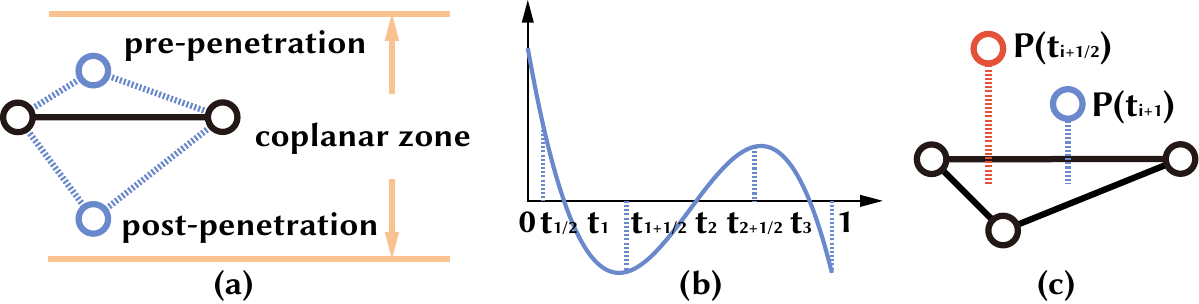}
    \caption{\textbf{Illustrations of The Conservative TOI Optimization.} (a) Both pre- and post-penetrations can be returned as valid coplanar solutions; (b) The reference frame for $t_{i+1}$ is $t_{i+1/2}=\left(t_{i}+t_{i+1}\right)/2$; (c) The distances of the reference frame and the target frame should have the same sign.}
    \label{fig:ccd_toi}
\end{figure}
In contrast to exterior-point methods, where a boolean value indicates contact, conservatively evaluated time-of-impact (TOI) is crucial for truncated step lengths. However, complete collision elimination is not assured with the TOI due to a floating-point algorithm's limitations (\autoref{fig:ccd-fail}). This is because the TOC computation falls below a small threshold, leading to TOI either before or after the penetration point (\autoref{fig:ccd_toi}a).
To tackle this, we optimize TOI calculation, employing signed distance for VF/EE instances only, as VV/VE cases lack signed distance. Within a time step, a cubic equation can produce up to three roots within $t\in\left[0,1\right]$, with $0$ and $1$ as candidates. We denote the conservative roots as $\left\{t_0=0,t_1,t_2,t_3,t_4=1\right\}$. Illustrating with the vertex-triangle scenario: since non-penetration can arise from either side of a triangle mesh, signed distance alone cannot confirm penetration. Hence, we introduce a reference frame $t_{i+1/2}=\left(t_{i}+t_{i+1}\right)/2$ for the target frame $t_{i+1}$ (\autoref{fig:ccd_toi}b). The distance of these frames should share the same sign (\autoref{fig:ccd_toi}c). If the signs differ, we conservatively backtrack $t_i$ by gradually decrementing it, typically by multiplying it by 0.9, until the signs match.

Our collision detection solution demonstrates robust performance across a diverse range of scenarios, as detailed in \autoref{sec:experiments}.

\section{Experiments and Results} \label{sec:experiments}

Our experiments and comparisons are conducted using a combination of CUDA and C++ on two desktop PCs configured as follows: (i) The first system is equipped with an Intel i9 13900K CPU with 24 cores and 128 GB of RAM, dedicated to IPC \cite{Li2020IPC}. (ii) The second system features an Intel Xeon Gold 6226R CPU with 16 cores, 128 GB of RAM, and an NVIDIA 4090 GPU with 24 GB of VRAM, which is used for GIPC \cite{gipc2024} and our implementations.
We use double precision as default for comparison purposes.
We emphasize that our single-precision version also demonstrates robust performance.

\subsection{Unit tests}
We perform a series of standard tests on Erleben's degenerated collision handling benchmarks \cite{erleben2018methodology}, scenarios involving sharp contact and stiff materials, and frictional contacts with large deformation. These tests serve to demonstrate the accuracy and reliability of our simulator.
\begin{figure}[h]
    \centering
    \includegraphics[width=\linewidth]{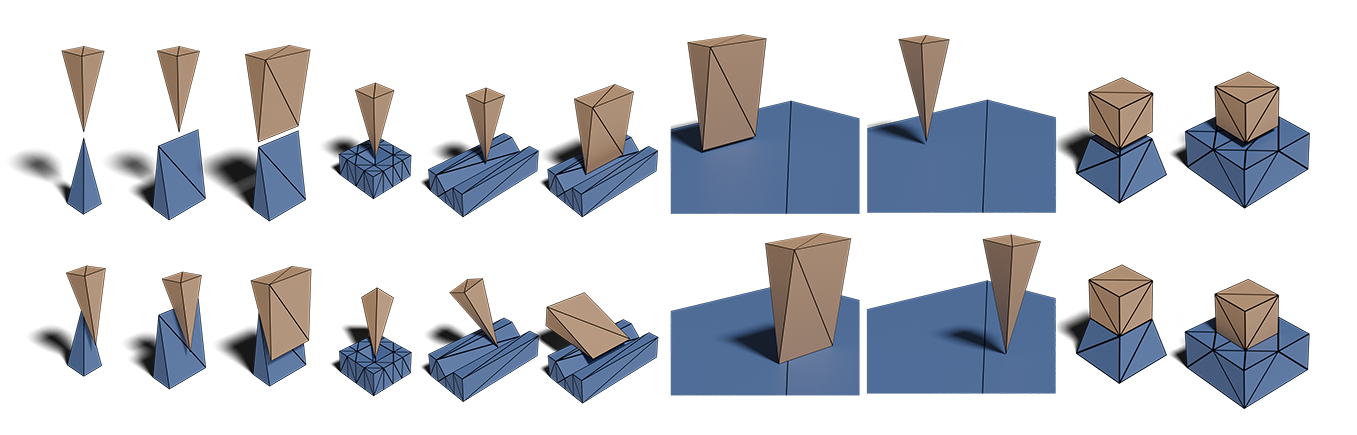}
    \caption{\textbf{Erleben's Tests.} Our method successfully passed the Erleben's tests in degenerated cases of mesh-based collision handling.}
    \label{fig:erleben}
\end{figure}

\paragraph{Erleben's test}
Our method robustly handled the degenerated collisions in Erleben's benchmarks with a challenging material setting as $E=10$ MPa and $\nu=0.4$, which is illustrated in \autoref{fig:erleben}.

\begin{figure}[h]
    \centering
    \includegraphics[width=0.9\linewidth]{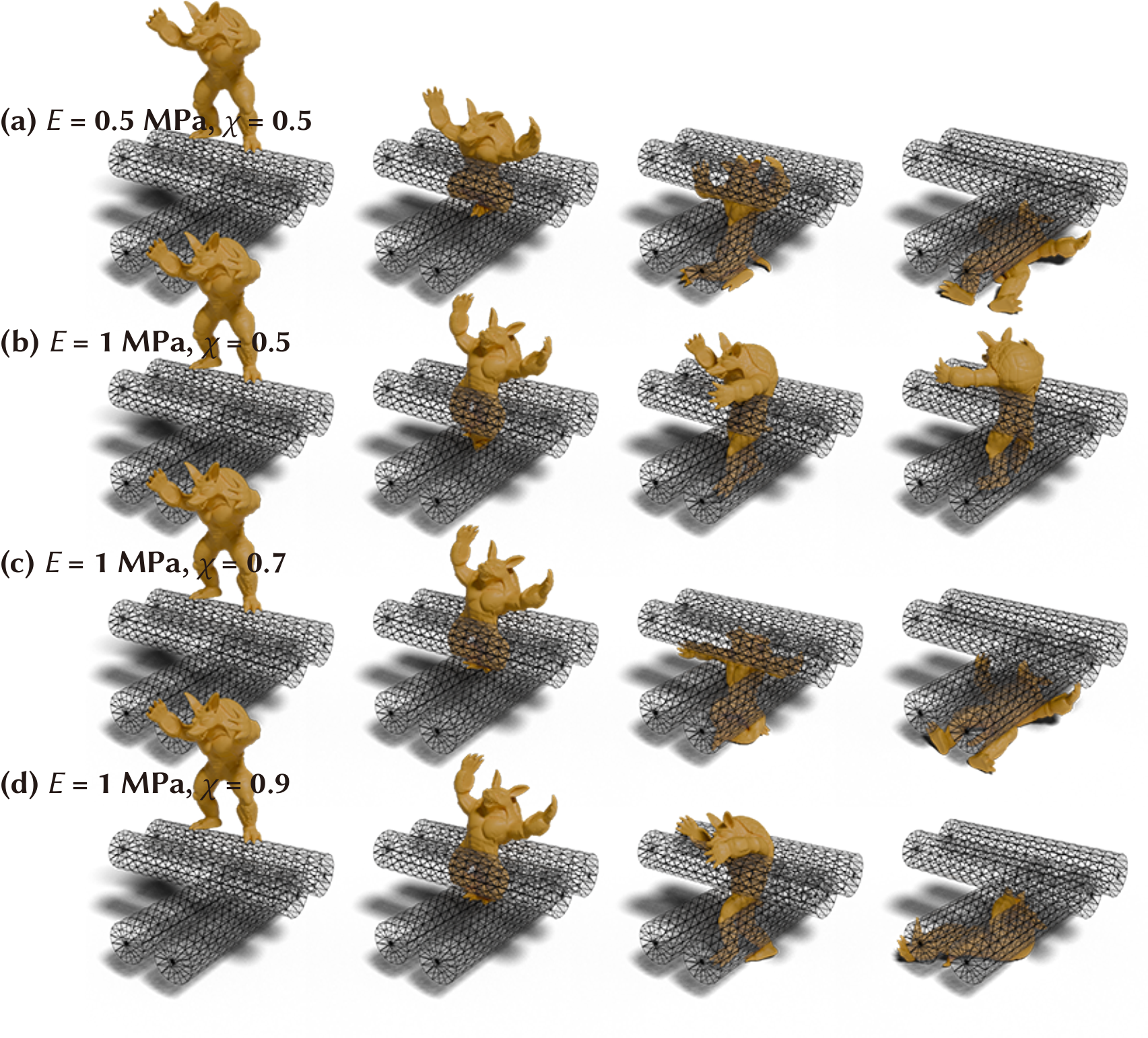}
    \caption{\textbf{Roller Test.} We employ an armadillo-rollers benchmark \cite{verschoor2019efficient} to conduct unit tests on friction. In the absence of friction, the armadillo is unable to roll down to the ground. However, introducing a friction coefficient of $\chi = 0.5$ enables the armadillo to roll smoothly down to the ground.
    }
    \label{fig:armadillo-roller}
\end{figure}



\paragraph{Frictional Contact with Large Deformation}
We assess the resilience of our approach in simulating frictional contact alongside large deformation through the armadillo-rollers benchmark. As depicted in \autoref{fig:armadillo-roller}, our fully implicit friction model adeptly captures nuanced distinctions across diverse parameter configurations. For instance, when Young's Modulus is set at $0.5$ MPa, an effective friction coefficient of $\chi = 0.5$ facilitates the armadillo's rolling motion onto the ground. Conversely, a Young's Modulus of $1$ MPa demands higher friction coefficients, such as $\chi = 0.7$ or $\chi = 0.9$, to ensure success during testing. Moreover, stick-slip transitions are effectively reproduced with a friction coefficient of $\chi = 0.9$ (see the supplemental video).
Across various Young's Moduli and friction settings, our method exhibits good scalability, as evidenced by the average per-frame costs of 12.5s, 9.9s, 21.1s, and 13s, respectively, from top to bottom.

\subsection{Validation of Scalability \& Correctness}

\begin{figure}[h]
    \centering
    \includegraphics[width=\linewidth]{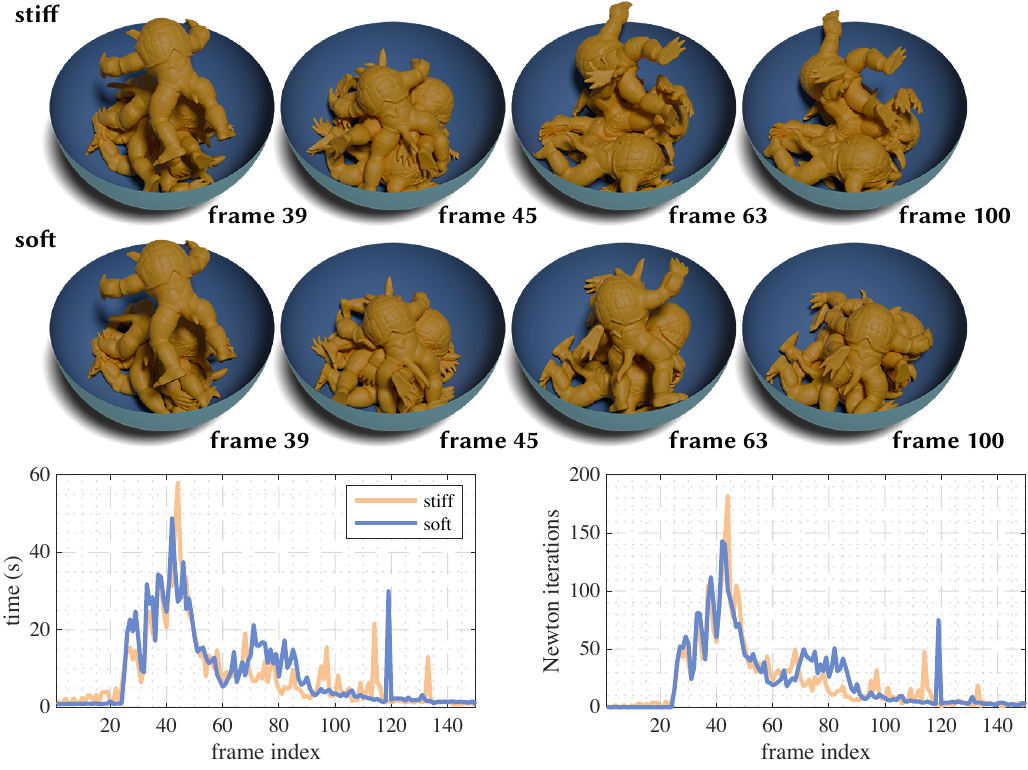}
    \caption{\textbf{Varying Young's Modulus.}
    Our method shows good scalability regarding the problems with different stiffness.
    }
    \label{fig:armadillo}
\end{figure}
\begin{figure}[h]
    \centering
    \includegraphics[width=\linewidth]{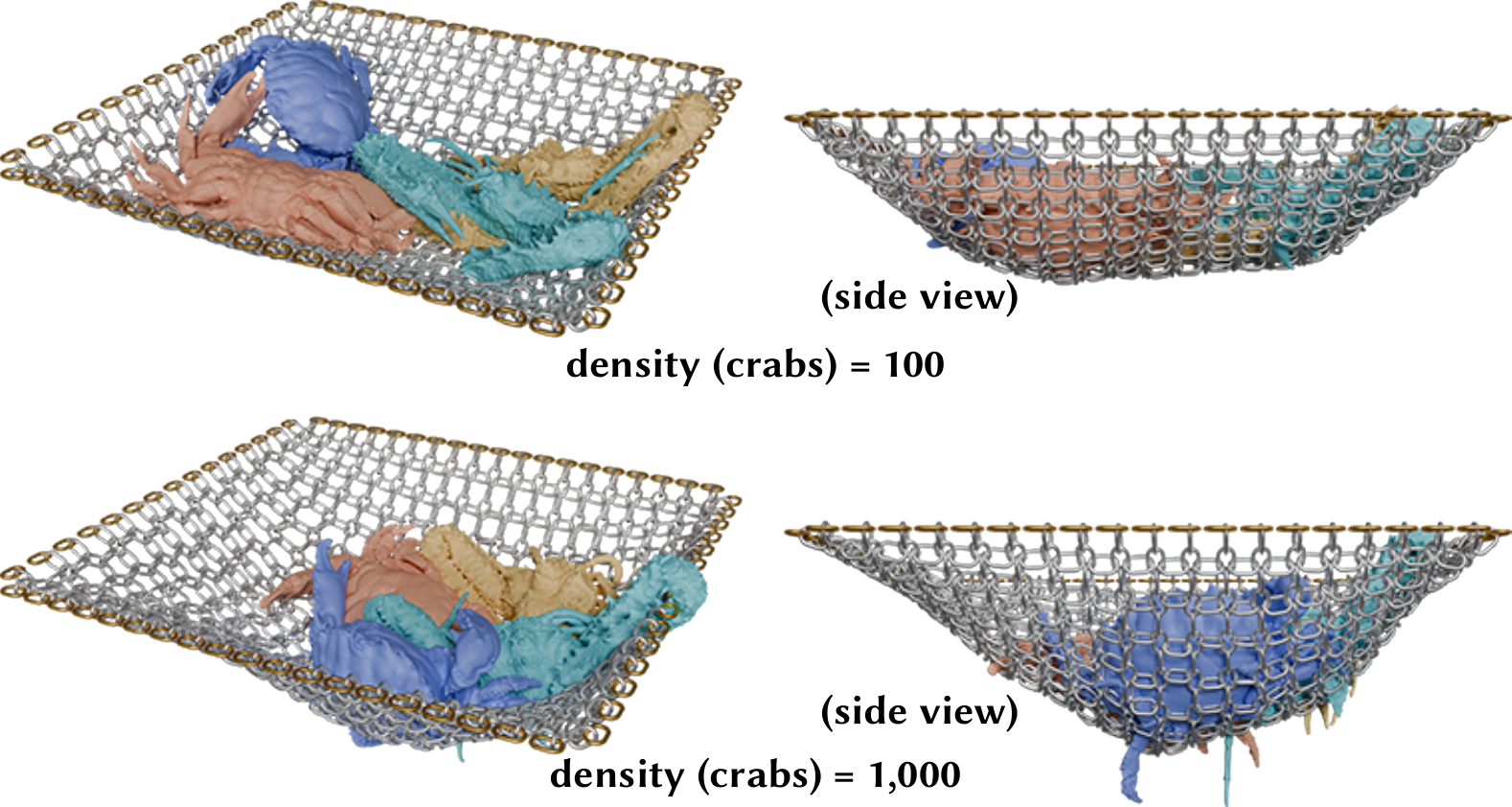}
    \caption{\textbf{Varying Densities.} Illustration showcasing two simulations involving four crabs descending onto a net. The net is modeled with a high stiffness of $E=100$ MPa, while the crabs are modeled with a lower stiffness of $E = 1$ MPa. To enhance visual realism, a reduced density of $100$ kg/m$^3$ is incorporated for the crabs (top) instead of the default density ($1,000$ kg/m$^3$, bottom), ensuring visually pleasing and realistic outcomes.
    }
    \label{fig:crabs}
\end{figure}

\paragraph{Varying Material Properties}
We explore the influence of Young's modulus ($E$) and density on the efficiency and visual effects of elastodynamic contact simulations.
In \autoref{fig:teaser}, we present a challenging experiment involving the dropping of four puffer balls onto chain-nets with varying material stiffness. Our approach effectively captures the complexities of this heterogeneous simulation, yielding controllable and realistic outcomes.
As illustrated in \autoref{fig:armadillo}, the Young's modulus does not emerge as the predominant factor influencing efficiency.
In this experiment, we use armadillos with varying stiffness levels—specifically 500 KPa and 1 MPa—arranged in a stack within a bowl for evaluation. The different Young's Moduli do not result in a noticeable difference in performance, as shown in the timing and Newton iterations plot in \autoref{fig:armadillo}.
In \autoref{fig:crabs}, we show two simulations involving four crabs falling onto a net. In this scenario, the net is characterized by a high stiffness of $E=100$ MPa, while the crabs are assigned a lower stiffness value of $E = 1$ MPa. However, the default density of $1,000$ kg/m$^3$ makes the net overly stretchy (bottom), while a reduced density of $100$ kg/m$^3$ for the crabs results in more rigid behaviors for the net (top). Our method demonstrates excellent scalability across different material properties, producing exceptional results.

\begin{figure}[h]
    \centering
    \includegraphics[width=\linewidth]{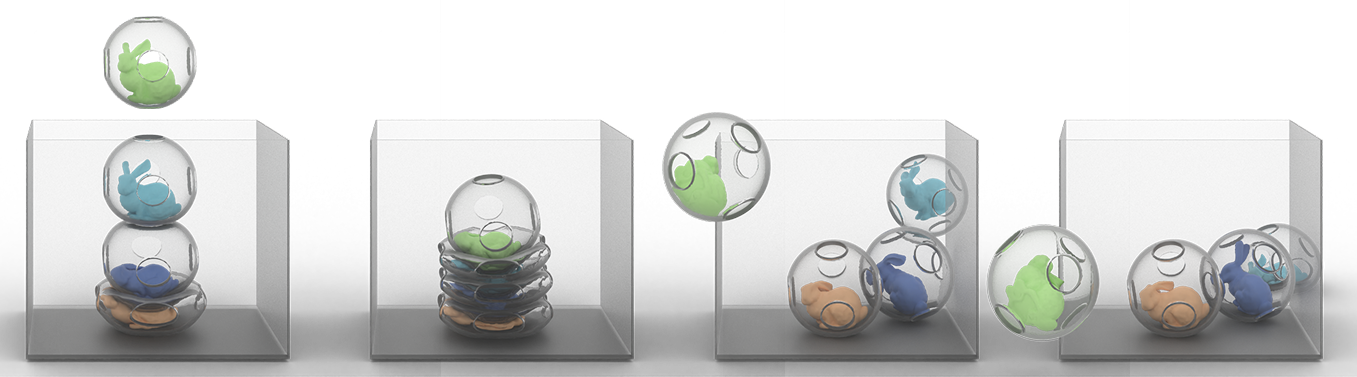}
    \caption{\textbf{Neo-Hookean Bunnies Encased in ARAP Balls.} In this illustration, we highlight the dynamic interplay between materials of varying stiffness—symbolized by the contrast between the soft bunnies with an elastic modulus $E=10$ KPa and the stiff elastic balls with $E=1$ MPa.
    }
    \label{fig:bunny-balls}
\end{figure}

\paragraph{Coupling between Different Elasticity Models}
\autoref{fig:bunny-balls} depicts soft Neo-Hookean bunnies ($E=10$ KPa) inside stiffer ARAP balls ($E=1$ MPa), showcasing the interaction between materials of contrasting stiffness. The bunnies and balls exhibit a strong coupling, highlighting the dynamic response due to material differences.

\begin{figure}[h]
    \centering
    \includegraphics[width=\linewidth]{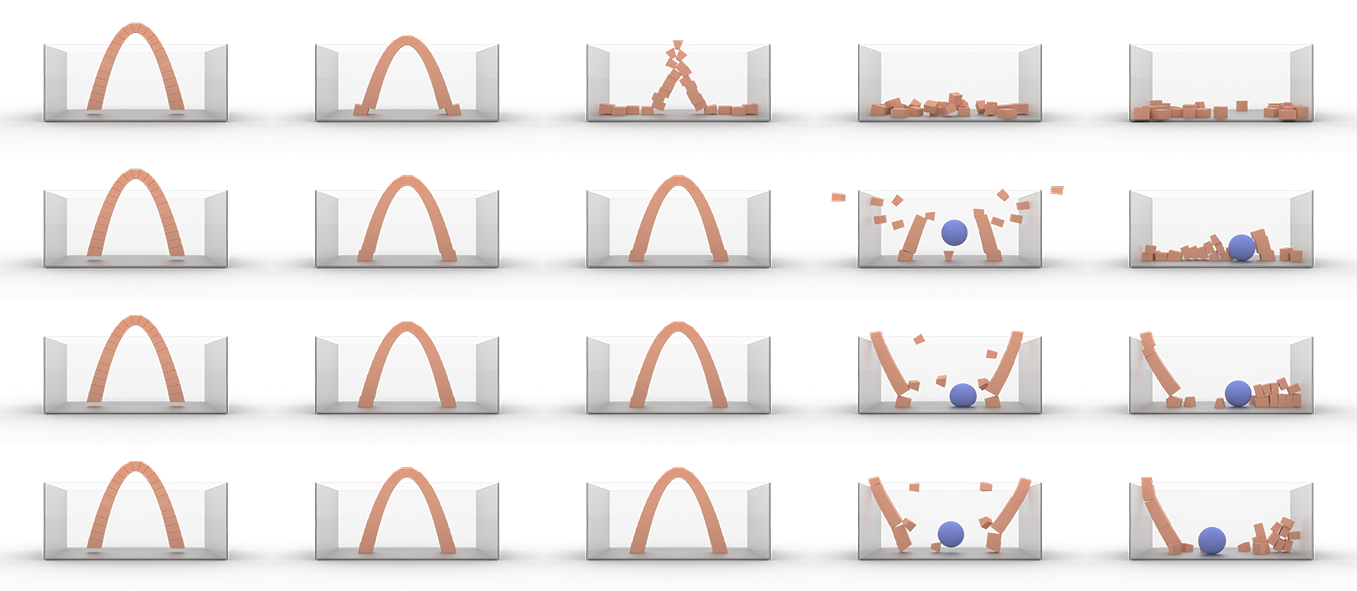}
    \caption{\textbf{Mansonry Archs.} The deformable arch ($E = 20$ GPa) attains static stable equilibrium through our frictional contact model. Once the arch stabilizes, a deformable ball ($E = 1$ MPa) is dropped onto it, causing the arch to collapse into the aquarium. Our method robustly handles frictional contact at a large time step size of $1/30$ s. ($\chi=0,0.1,0.5,0.9$ from top to bottom)}
    \label{fig:arch}
\end{figure}

\begin{figure}[h]
    \centering
    \includegraphics[width=\linewidth]{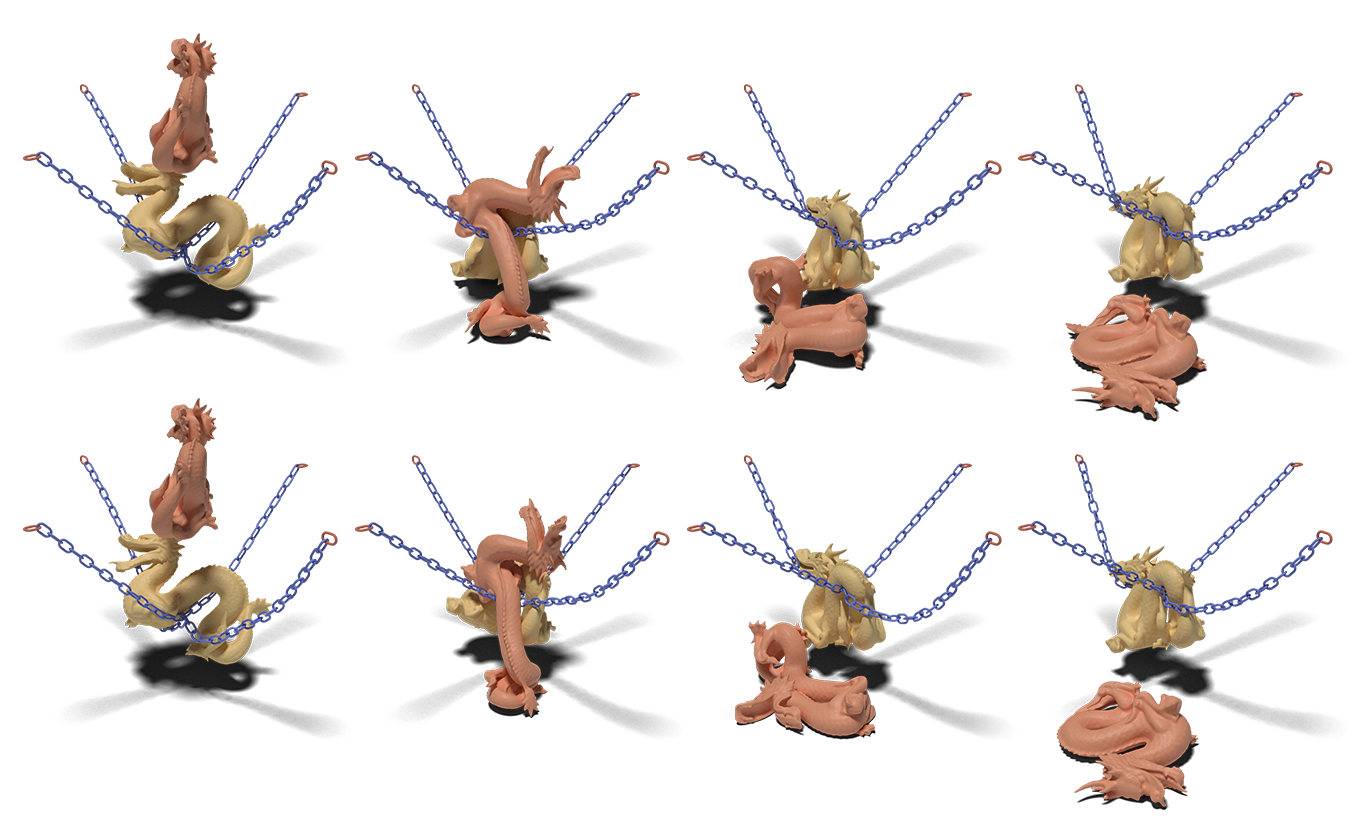}
    \caption{\textbf{Varying Resolution.} Different resolutions are applied to the dragon models for simulations under identical scenarios. The simulation with 65 K tetrahedras (upper) showcases comparable deformations across frames when compared to the simulation with 320 K tetrahedras (lower). Statistical analyses additionally validate the similarity in shapes between the two simulations.
    }
    \label{fig:tessellation}
\end{figure}

\begin{figure}[h]
    \centering
    \includegraphics[width=\linewidth]{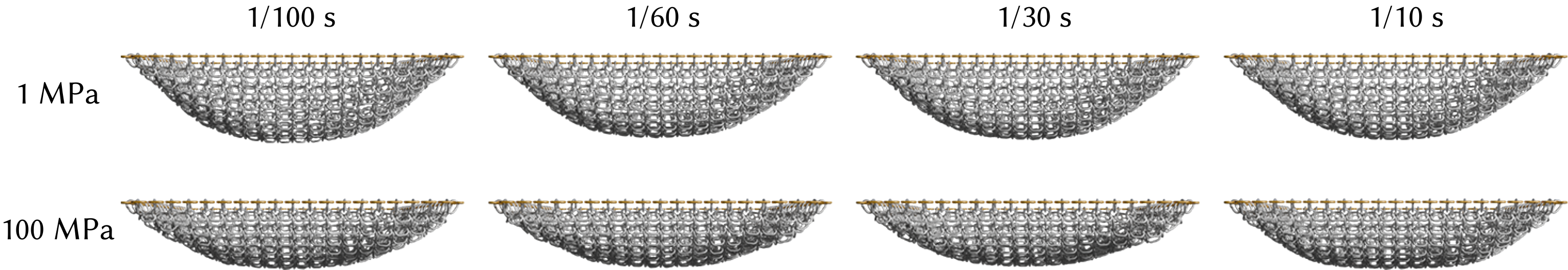}
    \caption{\textbf{Varying Time Step Sizes.} This figure depicts simulations of a structure with differing time steps (1/100 s to 1/10 s) and Young's modulus values (1 MPa, 100 MPa). It illustrates consistent equilibrium states across time steps, with a caution that larger steps may induce numerical damping. For dynamic visualizations, see the supplemental video.}
    \label{fig:time-step}
\end{figure}

\begin{figure}[h]
    \centering
    \includegraphics[width=\linewidth]{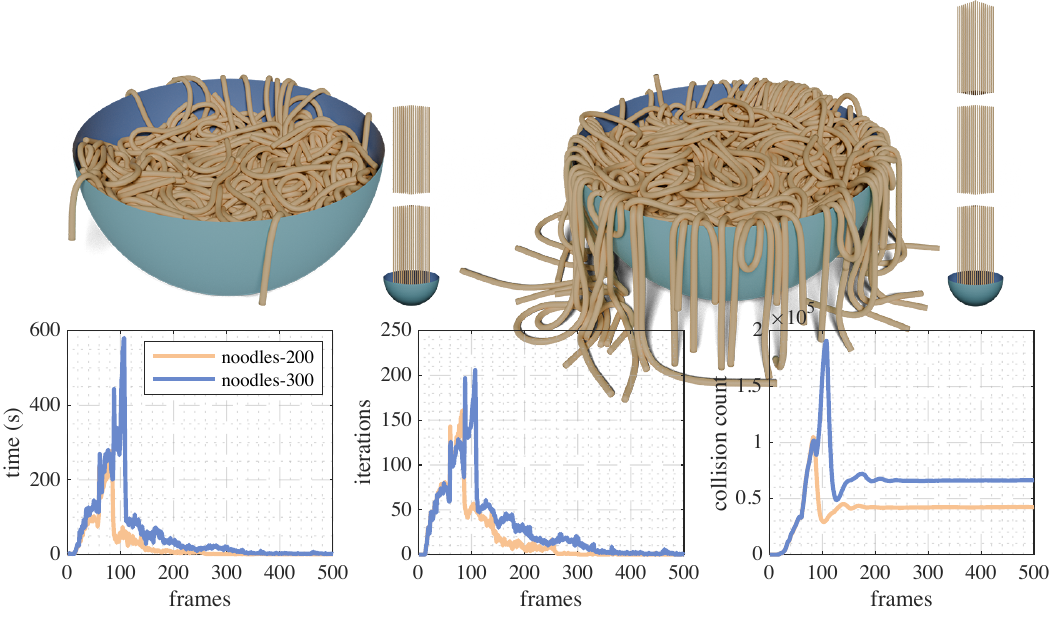}
    \caption{\textbf{Scalability.} This example illustrates the performance of a simulation method as the problem size scales from \emph{noodles-200} to \emph{noodles-300} (top row). It quantifies the scalability in terms of collision count, computational time, and iteration requirements per frame (bottom row). The moderate increase in time and iterations suggests that the method is capable of handling larger problem sizes with reasonable scalability.}
    \label{fig:noodles-300}
\end{figure}

\paragraph{Friction}
Our friction model can be precisely regulated through the coefficient $\chi$. In \autoref{fig:arch}, we successfully stack the masonry arch using $\chi = 0.1,0.5,0.9$. To provide a comparison with frictional contact, the frictionless scenario is illustrated in the top row of \autoref{fig:arch}.

\paragraph{Varying Resolution}
In the production phase, simulations are often previewed at lower resolutions. The critical consideration is whether simulations at lower resolutions can accurately reproduce results comparable to those obtained at higher resolutions. As demonstrated in \autoref{fig:tessellation}, our method effectively achieves this in the context of a scene depicting dragons dropping onto links.

\paragraph{Varying Time Step Sizes}
\autoref{fig:time-step} showcases simulations of a structure's response to different temporal resolutions and material stiffnesses, using time steps ranging from 1/100 s to 1/10 s and Young's modulus values of 1 MPa and 100 MPa. The uniform equilibrium states across various time steps suggest that the structure's response is relatively insensitive to the rate of loading, emphasizing the dominance of material properties and structural geometry in determining behavior.
However, the simulations also highlight a cautionary note on numerical damping, a computational artifact more pronounced at larger time steps that can obscure the true dynamic response of the structure. Therefore, while the simulations offer valuable insights into the material behavior under different conditions, the potential for numerical errors necessitates careful interpretation of these results. The supplemental video serves as a crucial resource for verifying the simulations by providing a real-time visualization of the structure's dynamics.

\begin{figure}[h]
    \centering
    \includegraphics[width=0.9\linewidth]{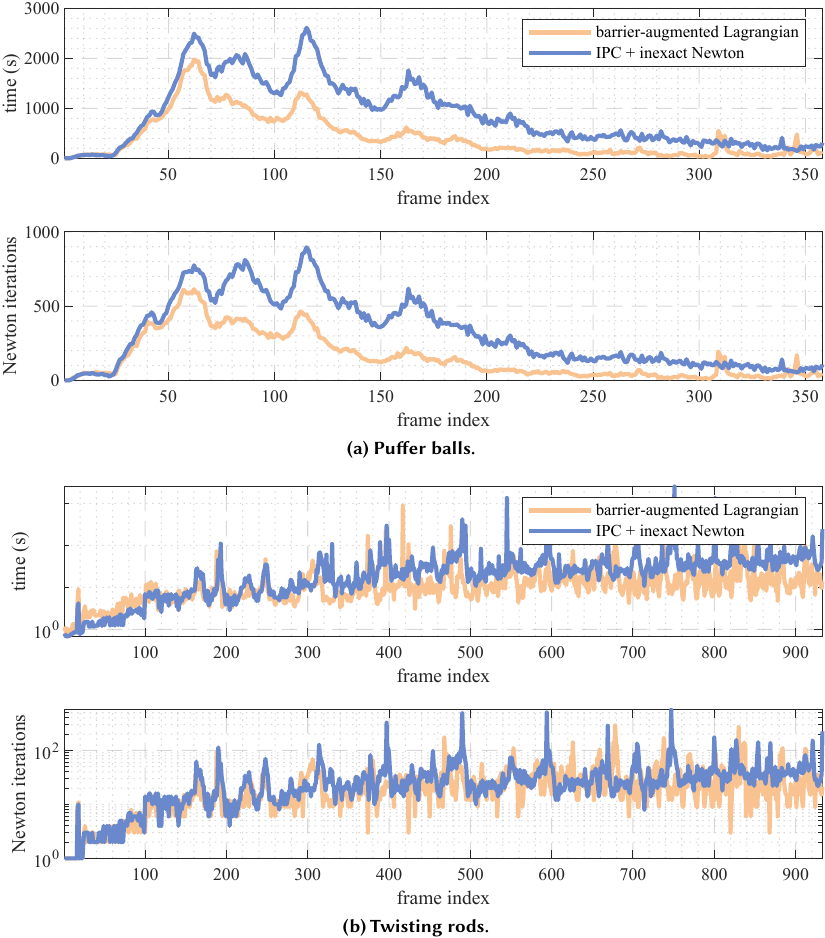}
    \caption{\textbf{Comparison with GPU-based Inexact Newton.} Our barrier-augmented Lagrangian demonstrates superior performance over inexact Newton, particularly in demanding scenarios characterized by intensive collisions. (a) \autoref{fig:teaser}b, (b) \autoref{fig:rods}.}
    \label{fig:auglag_vs_newton}
\end{figure}

\begin{figure}[h]
    \centering
    \includegraphics[width=0.9\linewidth]{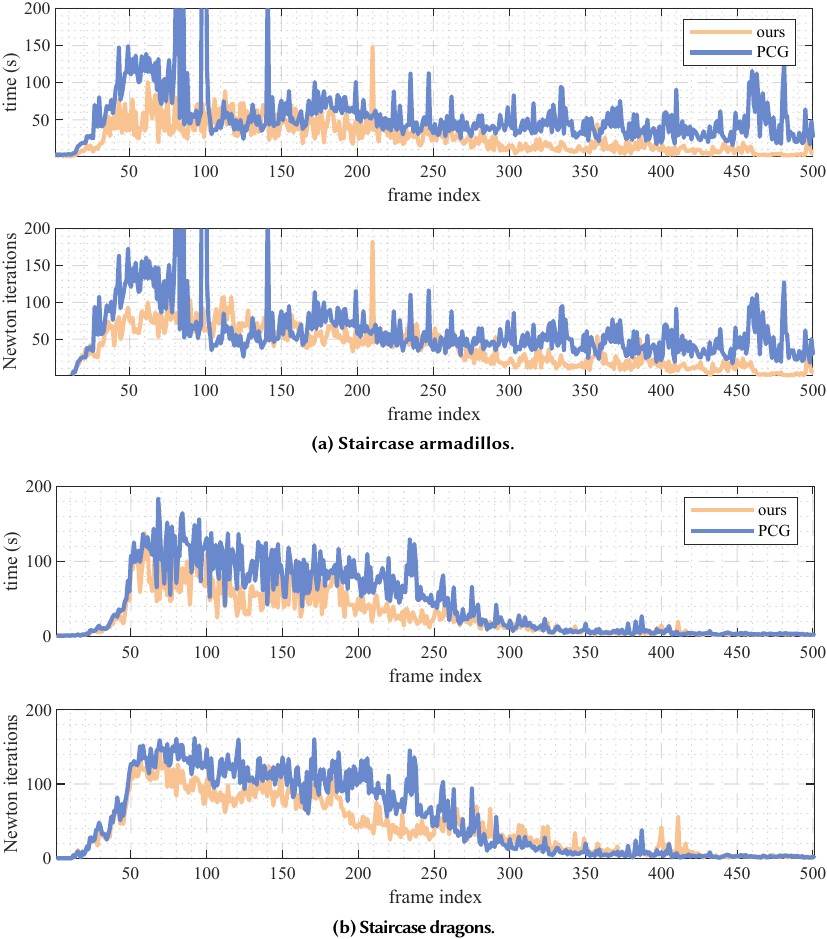}
    \caption{\textbf{Comparison with GPU-based PCG.} Our block-Jacobi warm starts significantly enhances the stability of solutions for ill-conditioned linear subproblems, as illustrated in the staircase scenarios shown in \autoref{fig:staircase}. Consequently, this results in overall performance improvements over traditional PCG methods.}
    \label{fig:pcg-cmp}
\end{figure}

\begin{figure}[h]
    \centering
    \includegraphics[width=0.9\linewidth]{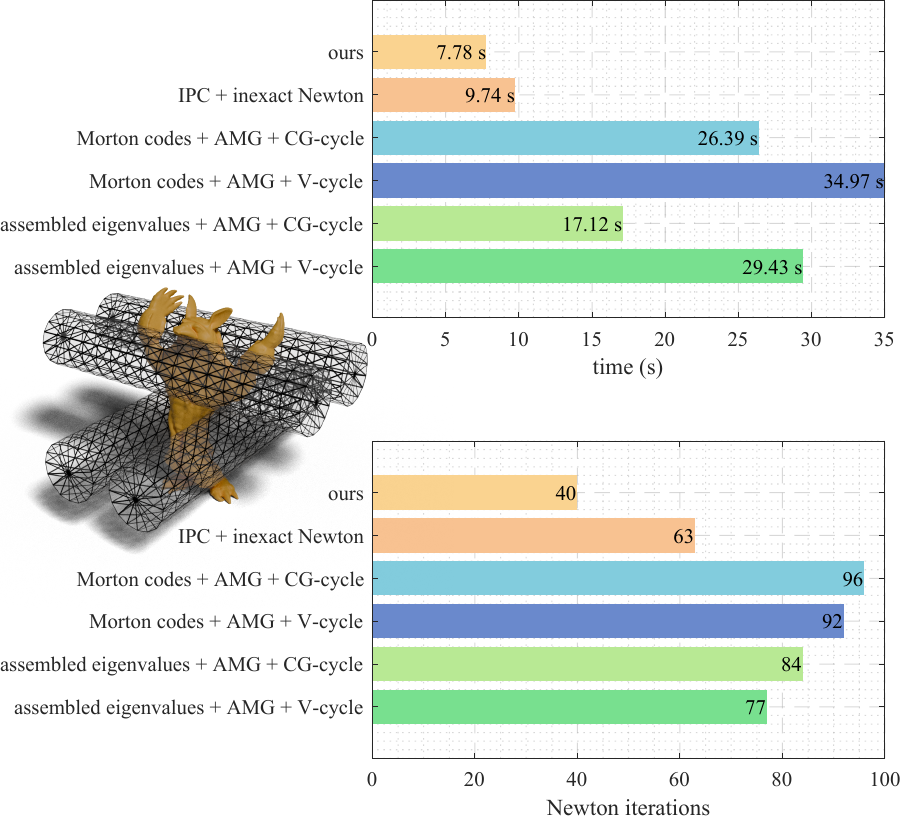}
    \caption{\textbf{Comparison with Morton-code Sorting and AMG.} This offers a comparative analysis of different approaches for solving the given problem (the roller test, \#cols = 6,682), highlighting the trade-offs between computational efficiency and convergence behavior.}
    \label{fig:amg}
\end{figure}

\begin{figure*}[h]
    \centering
    \includegraphics[width=\linewidth]{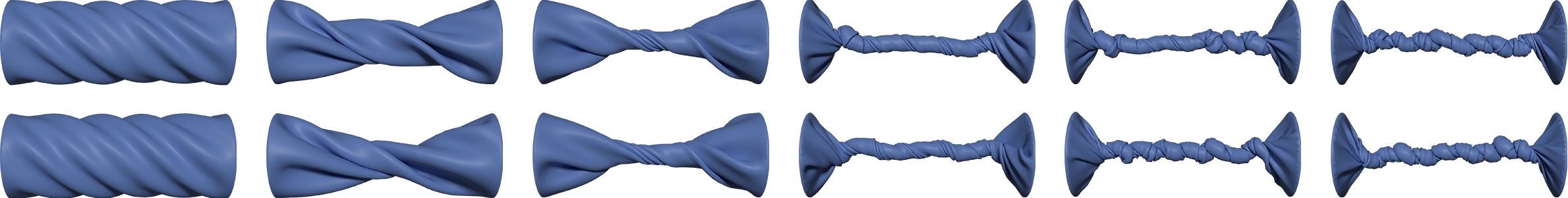}
    \caption{\textbf{Twisting a Cylindrical Mat.} The sequence displays side-by-side results of two methods applied to twist a cylindrical mat: our proposed method (top row) and IPC (bottom row). Both techniques achieve plausible outcomes. On average, our method completed each step in just 5.7 seconds, which is 19.3 times faster than IPC, demonstrating a significant improvement in processing speed without compromising the quality of the results.}
    \label{fig:twisting-cylinder}
\end{figure*}

\paragraph{Scalability}
To evaluate scalability, we compare the simulation of 200 and 300 noodles, respectively (\autoref{fig:noodles-300}). The corresponding increase in time and iterations per frame with the enhanced problem size is moderate, indicating that the method scales very well. This slight increase in resource demand suggests a robust algorithm capable of accommodating larger simulation parameters without a significant loss in efficiency.

\subsection{Ablation Study}
\paragraph{Barrier-Augmented Lagrangian}
As depicted in \autoref{fig:auglag_vs_newton}, statistical analysis of both puffer balls and twisting rod scenarios demonstrates significant improvements in our barrier-augmented Lagrangian method over the original IPC method with an inexact Newton solver. Specifically, our method achieves a $2.03\times$ speedup compared to the inexact Newton method, along with a $2.01\times$ enhancement in convergence for the puffer balls scenario. Similarly, in the case of the twisting rod, we observe a $2.3\times$ speedup accompanied by a $1.3\times$ improvement in convergence.
It is also noteworthy that the inexact Newton method encounters a convergence issue in the twisting-rods scenario at frame 933, while our barrier-augmented Lagrangian method does not have any problems (see \autoref{fig:rods}).

\paragraph{Block-Jacobi Warm Start}
In \autoref{fig:pcg-cmp}, we present a detailed comparison between our innovative block-Jacobi warm start technique and the traditional PCG method. Our approach showcases significant improvements in both computational efficiency and convergence performance. Specifically, our method demonstrates notable speedups, achieving overall performances of $2.08\times$ and $1.53\times$ faster than GPU-optimized PCG, in the respective staircase scenarios. This performance is particularly noteworthy considering that PCG serves as a strong baseline with our scalable storage formats and SpMVs, especially in scenarios where collision constraints vary from iteration to iteration. These results underscore the effectiveness of our warm start approach in efficiently navigating through challenging problem spaces characterized by poorly tessellated meshes.

\paragraph{Morton-code Sorting with Modified AMGs}
Node sorting alone typically does not inherently improve the convergence of iterative solvers like PCG. The convergence of PCG is primarily influenced by the eigenvalue distribution of the preconditioned matrix rather than its bandwidth or sparsity pattern alone. Therefore, for a fair comparison, we integrate node sorting with an algebraic multigrid (AMG). In this approach, presmoothing involves an accelerated Jacobi iteration utilizing Chebyshev polynomials \cite{wang2015chebyshev}, and the restriction-prolongation operations follow a similar methodology as described in \cite{wu2022gpu}. At the coarsest level (the fourth level), featuring diagonal blocks of size $96\times96$ (with at most one remainder block whose size is less than $96\times96$), we employ either a PCG (CG-cycle) or Cholesky factorization (V-cycle).
As depicted in \autoref{fig:amg}, our node sorting method based on assembled eigenvalues demonstrates improved convergence compared to Morton code sorting. Although the V-cycle incurs a higher computational cost than the CG-cycle, its convergence speed remains comparable. This is because achieving solutions with higher accuracy in linear systems can lead to unnecessary computational overhead. Furthermore, using AMG does not improve convergence in this case, as the dominant errors persist as high-frequency errors, which aligns with our expectations.

\begin{table*}[h]
  \centering
  \caption{\textbf{Statistics for Testing Scenarios.} This table details the total numbers of tetrahedra (\#tets), Degrees of Freedom (\#DOFs), and surface triangles (\#tris). Key simulation parameters such as time step ($h$), material density, Young's Modulus ($E$), Poisson Ratio ($\nu$), collision offset ($\hat{d}$), and frictional coefficient ($\chi$) are provided. Additionally, the table includes both average and maximum numbers of constraints (\#cons), the total number of Newton iterations per step, the average computational cost per step, and the comparative speedup achieved against IPC. Note that we simply use the same value for the friction mollification threshold $\epsilon_v$ and $\hat{d}$.}
  \small
  \begin{tabular}{c|c|c|c|c|c|c|c|c|c}
    \toprule
    Scenario & \#tets / \#DOFs / \#tris & $h$ (s) & \begin{tabular}[c]{@{}c@{}} density (kg/m$^3$), \\ $E$ (Pa), $\nu$\end{tabular} & $\hat{d}$, $\epsilon_v$ & $\chi$ & \begin{tabular}[c]{@{}c@{}} \#cons \\ (avg. / $\max$)\end{tabular} & \begin{tabular}[c]{@{}c@{}}avg. \#iters\\ (Newton)\end{tabular} & \begin{tabular}[c]{@{}c@{}} avg. cost\\ per-step (s)\end{tabular} & \begin{tabular}[c]{@{}c@{}}speedup\\ vs. IPC\end{tabular}\\
    \midrule
    Puffer Balls on Nets & 1.76M / 801K / 1.6M & 1/30 & 1e3, 5e5 / 1e9, 0.4 & 1e-3 & 0.3 & 228K / 292K & 156.8 & 427 & $80.1\times$\\
    \hline
    Dragons-Pachinko & 1.49M / 379K / 773K & 1/30 & 1e3, \begin{tabular}[c]{@{}c@{}}5e5 ($\times 2$)/\\1e6 ($\times 3$)\end{tabular}, 0.4 & 1e-3 & 0.3 & 4.9K / 18K & 41.4 & 29.1 & $73.9\times$\\
    \hline
    Staircase-Armadillos & 300K / 94K / 187K & 1/30 & 1e3, 7.5e5, 0.4 & 1e-3 & 0.5 & 3.2K / 3.2K & 38 & 26.7 & $47.2\times$\\
    \hline
    Staircase-Dragons & 376K / 120K / 240K & 1/30 & 1e3, 7.5e5, 0.4 & 1e-3 & 0.5 & 3K / 5.4K & 41.9 & 28.5 & $52\times$\\
    \hline
    Roller Test & 100K / 31K / 62K & 1/30 & 1e3, 1e6, 0.4 & 1e-3 & 0.9 & 1.6K / 5.8K & 35.4 & 12.5 & $31.4\times$\\
    \hline
    Armadillos \& Bowl & 826K / 192K / 238K & 1/30 & 1e3, 5e5, 0.4 & 1e-3 & 0.1 & 2.2K / 9.7K & 8.2 & 3.4 & $60.3\times$ \\
    \hline
    \begin{tabular}[c]{@{}c@{}}Crabs on Nets\\(light crabs)\end{tabular}
    & 2.2M / 810K / 1.2M & 1/30 & 1e2 / 1e3, 5e5, 0.4 & 1e-3 & 0.3 & 32K / 52K & 34.5 & 48.8 & $77.5\times$ \\ 
    \hline
    Twisting Rods & 355K / 70.4K / 51.6K & 1/30 & 1e3, 1e7, 0.4 & 1e-3 & 0 & 617K / 5.7M & 24.1 & 15.54 & $42.1\times$ \\
    \hline
    \begin{tabular}[c]{@{}c@{}}Twisting\\ Cylindrical Mat\end{tabular} & 64K / 20.9K / 41.8K & 1/30 & 1e3, 1e7, 0.4 & 1e-3 & 0 & 60K / 147K & 18.8 & 5.7 & $19.3\times$ \\
    \hline
    Noodles-200 & 934K / 375K / 749K & 1/30 & 1e3, 5e5, 0.4 & 1e-3 & 0.3 & 48.9K / 146.3K & 39.7 & 49.5 & $53.1\times$ \\
    \hline
    Noodles-300 & 1.4M / 562K / 1.1M & 1/30 & 1e3, 5e5, 0.4 & 1e-3 & 0.3 & 132.1K / 276K & 60.9 & 109.6 & $81.7\times$ \\
    \hline
    T-rex $\times 60$ & 9M / 2.2M / 2.9M & 1/30 & 1e3, 5e5, 0.4 & 1e-3 & 0.3 & 100.5K / 308.4K & 25.6 & 183.4 & \texttt{N/A} \\
    \bottomrule
  \end{tabular}
  \label{tab:stats}
\end{table*}

\subsection{Comparisons}
\paragraph{IPC \cite{Li2020IPC}}
We compare with the original IPC, making sure it utilizes full parallelization on the CPU by compiling CHOLMOD with Intel MKL and run the simulation on an Intel Core i9 13900K processor (24 cores), enabling a 24-thread Cholesky factorization for solving the linear systems.
\autoref{fig:twisting-cylinder} illustrates the effectiveness of two different computational methods in simulating the twisting of a cylindrical mat. Both methods produce visually comparable results; however, our method significantly outperforms IPC in computational efficiency, processing steps 19.3$\times$ faster on average. 
The demonstrated efficiency indicates that our method could provide considerable benefits to industries requiring fast and accurate simulations.
\autoref{tab:stats} showcases the statistics and quantifies the speedup achieved in representative cases relative to IPC.

\paragraph{Second-Order Stencil Descent \cite{lan2023second}}
In the study by \citet{lan2023second}, a novel GPU-accelerated algorithm is introduced for FEM elastodynamic simulations, leveraging interior-point methods to effectively handle complex scenarios involving extensive contact and collisions. This algorithm is notable for its use of complementary coloring and a hybrid sweep approach, which are well-suited for such applications. Nonetheless, these strategies may not fully address the specific challenges posed by stiff problems, such as significantly large stress resulting from challenging boundary conditions as in the simulation of twisting rods (\autoref{fig:rods}).
This example underscores our method's capability by stress testing four stiff rods with a Young's modulus of 10 MPa. These rods are subject to high-speed torsion from both ends, achieving an angular velocity of 5/12 revolutions per second over 18 complete turns. The image captures the deformation pattern, reflecting the rods' structural integrity and the material's resistance to the applied forces. Our method demonstrates proficiency in handling such demanding tests with large time steps, ensuring accurate results and computational efficiency.

\begin{figure}
    \centering
    \includegraphics[width=0.9\linewidth]{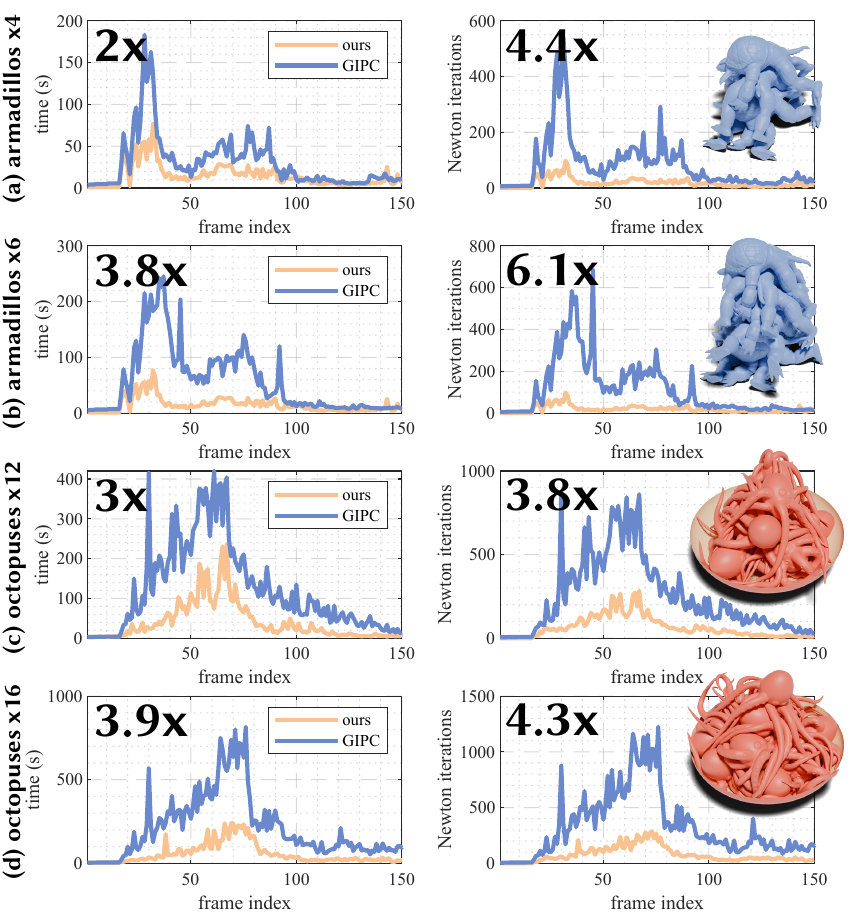}
    \caption{\textbf{Comparison with GIPC.} Our method surpasses GIPC in terms of both convergence speeds and rates across four benchmarks of varying scales.}
    \label{fig:gipc}
\end{figure}

\paragraph{GIPC \cite{gipc2024}}
The concurrent development of another GPU-based IPC method, termed \emph{GIPC}, employs a Gauss-Newton approximation for the contact Hessian matrix. This method solves the IPC system without the need for numerical eigendecompositions, an operation that is not easy to parallelize on the GPU. In contrast, our approach focuses on reformulating the nonlinear problem to make it easier to solve for both Newton's method and CG solvers.
In the comparative tests (see \autoref{fig:gipc}), we used simulations of stacked armadillos and octopuses with frictional contacts (where $\chi=0.5$) and aligned the Newton tolerance for both methods. Our method consistently outperforms GIPC, achieving up to $3.8\times$ in speedup and $6.1\times$ in Newton convergence. 
Specifically, GIPC encounters challenges in large-scale simulations due to suboptimal convergence speeds. While GIPC uses Newton-PCG for optimization, its performance can still be significantly affected by the conditioning of the system. The Multilevel Additive Schwarz (MAS) preconditioner utilized in GIPC effectively smooths out low-frequency errors commonly found in hyperelastic materials but struggles with the high-frequency errors that are typical in scenarios involving frictional contacts, leading to difficulties in larger-scale frictional contact simulations.

\section{Concluding Remarks and Discussion} 
In conclusion, this paper presents a GPU-optimized iterative method for robust and accurate simulation of elastodynamics and contact. Our barrier-augmented Lagrangian method, with the introduction of a slack variable, marks an improvement in the system conditioning and convergence speed of the primal-dual optimization process. The innovative GPU-based inexact Newton-PCG solver, enhanced by our novel subdomain corrections with early terminations, sets a new benchmark in performance, particularly for fully-implicit friction problems. Our scalable GPU strategies for Sparse Matrix-Vector Multiplication and Continuous Collision Detection exemplify the efficient handling of complex, large-scale simulations. The extensive experiments and ablation studies corroborate the robustness and efficiency of our method, showcasing its superiority in handling challenging scenarios involving frictional contact and nonlinear deformable solids with diverse material properties and time step sizes. Our method not only achieves significant speed improvements compared to existing IPC methods but also opens new frontiers in the practical application of GPU-based iterative methods to complex elastodynamic problems.

Looking ahead, there are several promising avenues for extending the work presented in this paper. One area of interest is the exploration of more advanced preconditioning techniques that could further enhance the efficiency and scalability of our GPU-based solver, especially for extremely large-scale simulations. Additionally, investigating the integration of machine learning algorithms to predict optimal solver parameters and improve real-time performance presents a novel research direction. Another potential development could involve adapting our methods to different types of physical simulations, such as fluid dynamics or coupled fluid-structure interaction problems, to broaden the applicability of our approach. Finally, a deeper exploration into the potential of hybrid CPU-GPU architectures for further optimization of computational resources and efficiency could be fruitful. These future endeavors could not only refine our current method but also open new horizons in the field of physically-based simulation.



\bibliographystyle{ACM-Reference-Format}
\bibliography{sample-base}

\appendix

\section{Additive Preconditioner (comparison only, not adopted)}
\label{app:as}
We adopt the standard additive preconditioner for comparison, in which the preconditioner $\mathbf{M}^{-1}=\sum\limits_{b}\mathbf{M}^{-1}_{b}$, where $\mathbf{M}^{-1}_{b}$ denotes a block-wise inverse with a size of $3N_b\times3N_b$, and $\mathbf{M}^{-1}$ the approximation of the matrix inverse $\mathbf{A}^{-1}$.
First, for a sparse linear system $\mathbf{Ax}=\mathbf{b}$, we have the general successive substitution scheme as
\begin{equation}    \mathbf{x}^{\left(k+1\right)}=\mathbf{x}^{\left(k\right)}+\mathbf{M}^{-1}\left(\mathbf{b}-\mathbf{A}\mathbf{x}^{\left(k\right)}\right),
    \nonumber
\end{equation}
where the superscripts $k$ denotes the iteration. We adopt the abbreviation $\mathbf{r}^{\left(k\right)}=\mathbf{b}-\mathbf{A}\mathbf{x}^{\left(k\right)}$.
The error correction (preconditioning) is defined as
\begin{equation}
\mathbf{x}^{\left(k+1\right)}=\mathbf{x}^{\left(k\right)}+\sum\limits_{i}\mathbf{B}_i^\intercal\left(\mathbf{B}_i\mathbf{A}\mathbf{B}_i^\intercal\right)^{-1}\mathbf{B}_i\mathbf{r}^{\left(k\right)},
    \nonumber
\end{equation}
where $\mathbf{B}_i$ is the $3N_b\times3N$ block-mapping matrix for each block $i$ that map the global system matrix to a block with predefined size $3N_b\times3N_b$.
The problem is equivalent as solving a subsystem
\begin{equation}    \left(\mathbf{B}_i\mathbf{A}\mathbf{B}_i^\intercal\right)\mathbf{e}_i=\mathbf{B}_i\mathbf{r}^{\left(k\right)}\mathbf{B}_i^\intercal
    \label{eq:linear-subs}
\end{equation}
for $\mathbf{e}_i$.
If $3N_b$ is sufficiently small, Problem \ref{eq:linear-subs} can be precomputed via matrix inversion, achievable through Gauss-Jordan elimination. In our implementation, each instance of Problem \ref{eq:linear-subs} at scales of $3\times3$, $9\times9$ and $27\times27$ is precomputed just once for a Newton step to establish preconditioning. Subsequently, corrections $\mathbf{e}$ at different scales are aggregated for preconditioning steps. For comparison, we employ a two-level additive preconditioner, requiring only two block-wise SpMVs on the GPU, thus optimizing efficiency.

\section{PCG Tolerance}
\label{app:tolerance}
Existing strategies, such as the truncated Newton method \\ (i) $\min\left(0.5,\sqrt{\left\|\nabla E\left(\mathbf{x}_{k}\right)\right\|_2}\right)\left\|\nabla E(\mathbf{x}_{k})\right\|_2$ \cite{nocedal2006numerical}, or leveraging condition numbers, for example, (ii) $u\kappa(\mathbf{A})\left\|\mathbf{x}_{k}\right\|_2$ \cite{golub2013matrix} or (iii) $u\kappa(\mathbf{A})\left\|\mathbf{b}\right\|_2$ \footnote{\url{https://scicomp.stackexchange.com/q/31024}}, where $u$ represents the machine epsilon, offer potential solutions but also come with their own set of drawbacks.
Specifically, the truncated Newton method often terminates the PCG solver too early, requiring extensive evaluation of expensive CCD and Hessian matrices, while the latter two approaches are too strict for mildly stiff cases. 

We instead apply a termination criteria based on relative residuals and check $\left\|\mathbf{r}^{k}\right\|_2\le10^{-4}\left\|\mathbf{r}^{0}\right\|_2$ to enable sensible early termination and enhance performance. 
If the system is ill-conditioned, this stopping criterion may be challenging to satisfy. Thus, we also monitor the residual's decrease over the most recent 100 PCG iterations. If it stops decreasing, we directly terminate the PCG and proceed to the line search. If the line search step size $\alpha$ is below $10^{-9}$, we return to the PCG method for an additional 100 iterations until $\alpha$ becomes larger.

\begin{figure}
    \centering
    \includegraphics[width=0.9\linewidth]{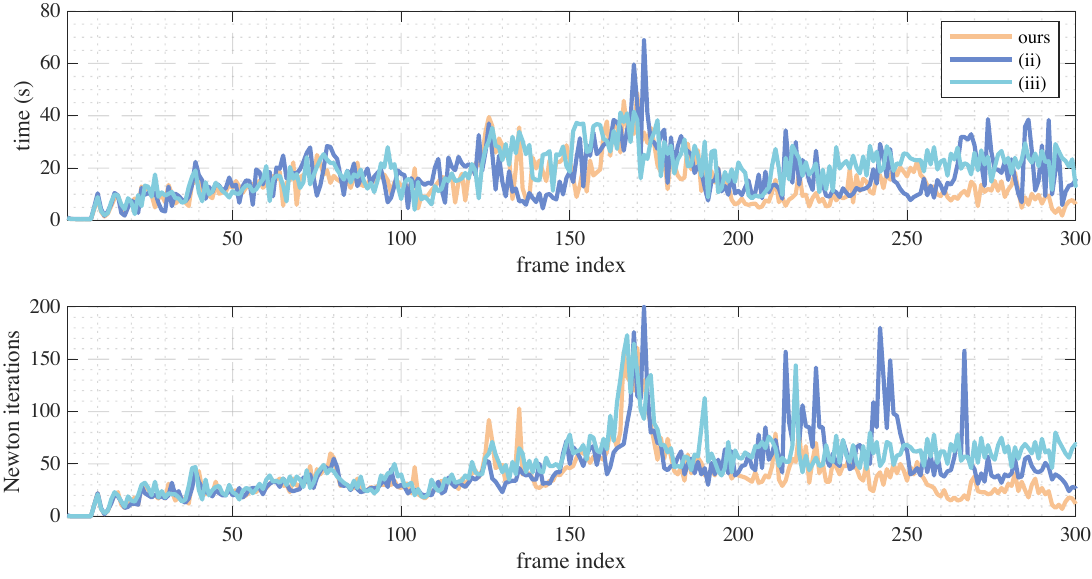}
    \caption{\textbf{Compare to PCG stopping criteria (ii) and (iii)} (scenario: roller test, $E=1$ MPa, $\chi=0.7$)).}
    \label{fig:tolerance}
\end{figure}

In the twisting-rods scenario (frame 933), criterion (i) failed to converge, resulting in a large number of collision pairs and a high residual ($>10^{15}$) after 5,000 iterations. By approximating the condition number in criteria (ii) and (iii) using our assembled eigenvalues across elasticity, collision stencils and diagonal mass matrix, they result in better convergence speed as shown in aggregated Newton iterations (8,882 and 9,871, respectively) compared to ours (10,657) (see \autoref{fig:tolerance}). However, criteria (ii) and (iii) were overly stringent on linear solves, resulting in higher aggregated timing costs (5,043s and 5,700s, respectively) compared to ours (4,287s).

\section{Sparse Matrix Data Structure}
\label{app:sp}
Below are the abstract declarations of our sparse matrix $\mathbb{L}$ \\(\texttt{sparse\_matrix\_L\_cuda}) and $\mathbb{C}$ (\texttt{sparse\_matrix\_C\_cuda}).
{
\footnotesize
\begin{lstlisting}
template<typename Real, unsigned int block_size = 3u>
class sparse_matrix_L_cuda
{
public:
    sparse_matrix_L_cuda(unsigned int dim, unsigned int nnz_lower);
protected:
    Real* dev_nonzeros_lower;
    unsigned int m_nnz_lower;
    unsigned int* dev_rows_to_cols;
    unsigned int* dev_offsets;
    unsigned int m_dim;
    unsigned int* dev_blocks_to_coords;
};
template<typename Real, unsigned int block_size = 3u>
class sparse_matrix_C_cuda
{
public:
    sparse_matrix_C_cuda(unsigned int nBlocks);
protected:
    Real* dev_nonzeros;
    unsigned int* dev_coords;
    unsigned int m_dim;
    unsigned int m_nBlocks;
    unsigned int* dev_blocks_to_coords;
};
\end{lstlisting}
}
To align with GPU data access, all nonzero entries are stored in device dense vectors (\texttt{dev\_nonzeros\_lower} and \texttt{dev\_nonzeros}) with corresponding offsets (\texttt{dev\_rows\_to\_cols}, \texttt{dev\_offsets}, and coordinates \texttt{dev\_coords}) to facilitate rapid access. Additionally, the auxiliary data \texttt{dev\_blocks\_to\_coords} provides a reverse mapping from blocks to coordinates, crucial for eliminating entries in sparse matrices, particularly for static boundaries.
\end{document}